\documentclass[prb,aps,singlecolumn,preprint,superscriptaddress]{revtex4-1}
\usepackage{bm}
\usepackage[colorlinks=true,linkcolor=blue,citecolor=blue]{hyperref}
\usepackage{times}
\usepackage{amsmath}
\usepackage{amssymb}
\usepackage{amsthm}
\usepackage{amsfonts}
\usepackage{enumerate}
\usepackage{latexsym}
\usepackage{ifpdf}
\usepackage{physics}
\newcommand{\beq}{\begin{equation}}
\newcommand{\eeq}{\end{equation}}
\usepackage{graphicx}
\usepackage{makeidx}
\usepackage{color}

\begin{document}

\title{Non-uniform magnetization profile in ferromagnetic heterostructures leading to topological Hall effect like signatures }

\author {Nandana Bhattacharya}
\affiliation{Department of Physics, Indian Institute of Science, Bengaluru 560012, India}
 \author {S. Middey}
\email{smiddey@iisc.ac.in }
\affiliation{Department of Physics, Indian Institute of Science, Bengaluru 560012, India}

\begin{abstract}

 Anomalous Hall effect (AHE), which arises when a current  is passed through a ferromagnetic material subjected to a perpendicular magnetic field,
 is proportional to the magnetization of the sample.
Additional hump-like features in AHE are often attributed to the presence of non trivial spin textures leading to topological Hall effect (THE). However, several recent reports have emphasized in context of ferromagnetic SrRuO$_3$ based heterostructures that the sample inhomogeneity can also result in THE-like features. In order to investigate this issue in general for any ferromagnetic heterostructure, we have considered a phenomenological model to calculate the changes in the shape of hysteresis loop due to various interfacial effects. These changes in the magnetization have been accounted for by considering that the interdomain magnetic coupling parameter ($\alpha$) varies exponentially with the distance from the interface along the growth direction of the heterostructure. In case of symmetric interfaces on both sides of a ferromagnet, we have considered the variation of $\alpha$ as a Gaussian function. We have found that the additional AHE contribution due to the net change in magnetization in such cases are akin to experimentally observed THE, even though we have not considered any topological quantity explicitly in our  model. Thus, we propose another situation with nonuniform magnetization profile that may be used to explain additional features in AHE, which might not necessarily be intrinsic THE.

\end{abstract}

\maketitle

 \section{Introduction}
Transition metal oxides (TMOs) are host of a plethora of collective phenomena such as magnetism, ferroelectricity, metal-insulator transition, unconventional superconductivity,  charge ordering, etc~\cite{Imada:1998p1039,Tokura:2006p797,Keimer:2015p179,Armitage:2010p2421}.   The tremendous advancement in thin film growth technologies over the last two decades have enabled the research community to  grow artificial structures of  these complex oxides with unit cell precision~\cite{Schlom:2008p2429,Zubko:2011p141,Hwang:2012p103,chakhalian:2014p1189,Bhattacharya:2014p65,Stemmer:2014p151,Middey:2016p305}.
Such heterostructuring~\cite{Bhattacharya:2014p65,Hellman:2017p025006,Chakhalian:2006p244,Chakhalian:2007p1114,Brinkman:2007p493,May:2009p892,Boris:2011p937,He:2012p197202,Gibert:2012p195,Grutter:2013p087202,Moon:2014p1,Hoffman:2016p041038,Ranjan:2020p041113,Yi:2016p6397,Matsuno:2016pe1600304} leads to subtle  modifications in  spin, charge,  orbital, and lattice sectors at the interface, resulting in a variety of emergent magnetic behaviors such as interfacial ferromagnetism, exchange bias, spin spiral magnetic phase, enhanced magnetic ordering temperature, topological Hall effect (THE), etc.   THE arises in materials with  non-zero scalar spin chirality [$\chi_{i,j,k}$= $\vb*{S_i}$.( $\vb*{S_j} \times \vb*{S_k}$),  $\vb*{S_i}$ denotes localized moment], which results in an internal fictitious magnetic field due to the real space Berry phase.  Since the first report of the observation of THE in MnSi~\cite{Neubauer:2009p186602}, THE has been demonstrated in a variety of systems having chiral spin configurations~\cite{Kanazawa:2011p156603,Surgers:2014p1,Kurumaji:2019p6456,Liu:2017p176809,Ojha:2020p2000021}.  As the simultaneous presence of broken inversion symmetry  and finite spin orbit coupling (either intrinsic or Rashba-type or both) can lead to a sizeable Dzyaloshinskii-Moriya (D-M) interaction, interfacial engineering of TMOs has  become a successful approach to achieve THE~\cite{Matsuno:2016pe1600304,Vistoli:2018p67,Shao:2019p182,Skoropata:2020peaaz3902}.

 Bulk SrRuO$_3$ (SRO) is a metallic ferromagnet with a magnetic transition temperature of 165 K~\cite{Cao:1997p321}.   The origin of THE-like signal in SRO-based heterostructures is under intense scrutiny in recent years~\cite{Matsuno:2016pe1600304,Ohuchi:2018p1,Wang:2018p1087,Qin:2019p1807008,Meng:2019p3169,Kan:2018p180408,Wu:2020p220406,Kimbell:2020p054414,Wang:2020p2468, Wysocki:2020p4.054402, groenendijk2020berry}.  While the origin of this phenomenon has been linked with the stabilization of nontrivial spin textures like skyrmions near the interface~\cite{Matsuno:2016pe1600304,Ohuchi:2018p1,Wang:2018p1087,Qin:2019p1807008,Meng:2019p3169}, another set of studies have emphasized that the sample inhomogeneity gives rise to THE-like features~\cite{Kan:2018p180408,Wu:2020p220406,Kimbell:2020p054414,Wang:2020p2468,Wysocki:2020p4.054402}. Since the shape of ferromagnetic hysteresis loop is generally affected due to the electronic, magnetic and structural changes across the interface~\cite{Alberca:2011p134402,Bruno:2011p147205},  we have considered another possible scenario with a nonuniform magnetization profile along the growth direction of the heterostructure~\cite{Singh:2012p077207}. We have examined whether  layer dependent hysteresis loops in such systems can lead to these THE-like signature without invoking any topological interpretation.  
 
 In this work, we have simulated the hysteresis loops using a phenomenological approach for a ferromagnet having layer dependent internal molecular field (we  refer  one unit cell along the growth direction as `layer' throughout the paper).  We have followed the mathematical model of Jiles et al. (Ref. ~\onlinecite{Jiles:1986p48}), which was developed to calculate the hysteresis loop for an isotropic ferromagnet.  In the simplest form, the effective magnetic field felt by an individual domain within a ferromagnet can be expressed as   $\vb*{H_e}=\vb*{H} + \alpha\vb*{M}$, where $\vb*{H}$ is the actual magnetic field within the domain   and $\vb*{M}$ represents the total sample magnetization per unit volume. The interdomain coupling is represented by a mean field parameter $\alpha$.
  It is assumed here that the change in the magnetization due to any interfacial effect can be accounted phenomenologically by  considering  $\alpha$ to be non-uniform and  strongly dependent  on  the distance ($z$) from the interface. For a ferromagnet, having an interface on one side,  hysteresis curves have been simulated considering that $\alpha$ varies exponentially with $z$ below a critical length scale $z_0$ and it becomes equal to $\alpha_0$ (mean field parameter of an uniformly magnetized ferromagnet) above that. To investigate effect of symmetric interfaces on both sides of a ferromagnet,  $\alpha$ has been considered to vary as a Gaussian function. We have found that the additional AHE, which is proportional  to the change in magnetization $\Delta M$  (defined as the change in magnitude of magnetization between the cases with constant and layer-dependent $\alpha$), appears very similar to what have been claimed experimentally as THE signature in several systems.~\cite{Liu:2018p122,Wang:2021p1108,Li:2019p21268,Meng:2019p3169,Yun:2018p034201,Shao:2019p182}.

\section{Details of simulation}

 Under the assumption of the presence of non-trivial spin textures like skyrmions ~\cite{Matsuno:2016pe1600304,Ohuchi:2018p1,Wang:2018p1087,Qin:2019p1807008,Meng:2019p3169}, the Hall resistivity ($\rho_{H}$) is expressed as
 \begin{equation}
 \rho_H =\rho_\mathrm{OHE} + \rho_{additional} = \rho_0H+  \rho_{additional}
 \end{equation}
 where $ \rho_{additional} $=$\rho_\mathrm{AHE}$+  $\rho_\mathrm{THE}$=$R_a M$ + $\rho_\mathrm{THE}$  with $\rho_\mathrm{OHE}$,  $\rho_\mathrm{AHE}$, $\rho_\mathrm{THE}$ denoting ordinary Hall effect (OHE), AHE and THE, respectively.  $\rho_0$, $R_a$ are constant and  $M$ is the magnetization.

 In case of a heterostructure with nonuniform magnetization along the growth direction, the total magnetization ($M$) can be considered as summation of two terms: $M'$ (magnetization of a uniform ferromagnet) and $\Delta M$ (this corresponds to the change in  $M'$  due to all interfacial effects). In such system, we can also express   $\rho_{additional}$ as a combination of two AHE contributions instead of considering  any intrinsic THE
 \begin{equation}
  \rho_{additional}= R_a M' + R_a \Delta M
  \end{equation}

   We have simulated hysteresis loops for $M$ and $M'$ using the method described in Ref.~\onlinecite{Jiles:1986p48}, which we  briefly discuss here.  The magnetization $\vb*{M}$ is  related to the $\vb*{H_e}$ and can be written as
   \begin{equation}\label{eq1}
  \vb*{M}=\vb*{M_s} f(H_e)
  \end{equation}
  where $f(H_e)$ is an arbitrary function with the following properties: $f \rightarrow$ 0 when $H_e$=0 and  $f \rightarrow$ 1 as $H_e \rightarrow  \infty $. $\vb*{M_s}$ is the saturation magnetization. Equation~\ref{eq1} does not consider any hindrance to the change in magnetization such as pinning of domain wall motion; therefore, represents only the anhysteretic magnetization curve ($\vb*{M_{an}}$) of a ferromagnet in reality.

  Thus, one may rewrite equation~\ref{eq1} as
    \begin{equation}\label{M1}
  {\vb*{M_{an}}(H_e)}=\vb*{M_s} f(H_e)
  \end{equation}

  Jiles et al.~\cite{Jiles:1986p48} further modelled $\vb*M_{an}$ by a modified Langevin function~\cite{Langevin:1905p203}
     \begin{equation}\label{M2}
  {\vb*{M_{an}}(H_e)}=\vb*{M_s} [coth(H_e/a)-(a/H_e)]
  \end{equation}
 The shape of the anhysteretic magnetization is determined by the parameter $a$, which has the same dimension as $H_e$.

  In presence of an external magnetic field, the size of ferromagnetic domains, which are oriented along the direction of the field, would grow in size. However,  pinning sites such as structural imperfections, impurity atoms, etc. disrupt domain wall motion and result in irreversible changes in magnetization curves.   The energy loss per unit volume due to the pinning can be represented as (see Ref.~\onlinecite{Jiles:1986p48} for the derivation)
   \begin{equation}
       E_{pin}(M)= K \int_{0}^{M}dM'
   \end{equation} where $K$ is a constant and  related to the average density of pinning sites,   the average pinning energy of all  sites for 180$^\circ$ oriented walls and the magnetization density of a domain aligned along the direction of the field.

   The magnetization energy is defined as the difference between the ideal, lossless case  $\int M_{an}(H_e)dB_e$ ($B_e$ is the effective magnetic induction) and the loss due to hysteresis $K\int dM$ i.e.

   \begin{equation}{\label{eq11}}
       \int M dB_e = \int M_{an}(H_e)dB_e - K\int\frac{dM}{dB_e}dB_e
   \end{equation}
   The corresponding differential form of the equation is
\begin{equation}{\label{eq12}}
    M=M_{an}-\beta K \frac{dM}{dB_e}
\end{equation}
where $\beta$ is +1 (-1) for the positive (negative) sweep of $H$ as the pinning process opposes the direction of the  applied field. This is the master equation that we have used to simulate the hysteresis curve. Since both sides of the equation~\ref{eq12} contain $M$, we have evaluated first a trial magnetization $M_1$ as a function of external magnetic field $H$ by the relation ${M_{1}(H)}={M_s} [coth(H/a)-(a/H)]$. The effective field is treated as $H_e=H+\alpha M_1$. We have used  equation ~\ref{M2} and ~\ref{eq12} sequentially to simulate hysteresis loops. Also, we have considered $H_e$ in Gaussian units, hence $B_e$ = $H_e$ and so $\frac{dM}{dB_e}$  can be written as $\frac{dM}{dH_e}$. The derivative of  $M$ with respect to $H_e$ has been calculated using the method of finite differences~\cite{Li:2017numerical}. The parameter $a$ is taken as 11 through out the paper.

 \section{Results and Discussions}

   \begin{figure}
\vspace{-0pt}
	\includegraphics[width=.48\textwidth] {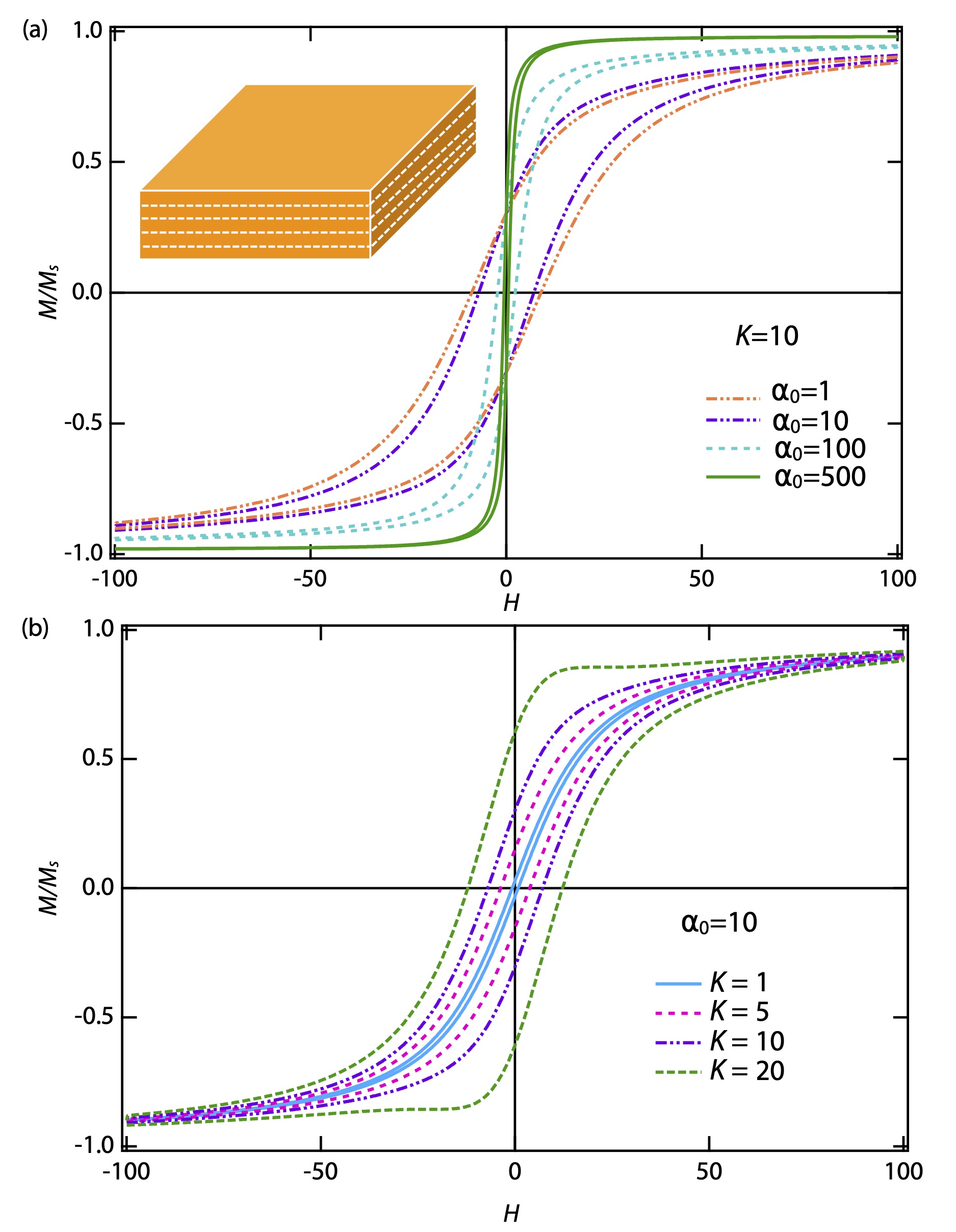}
	\caption{\label{Fig2} { Simulated hysteresis curves of a uniform  ferromagnet (without any interface):  (a) for different values of $\alpha_0$ keeping $K$=10, (b) for different values of $K$ with $\alpha_0$=10.}}
\end{figure}

  We first discuss the results of hysteresis loop simulation for a ferromagnet without any interfacial effects [see inset of Fig.~\ref{Fig2}(a)]. For this,  the material has been considered to be consisting of $N$=100 unit cells along the growth direction.  The total magnetization has been evaluated by summing the magnetization of each individual unit cell, characterized by a same constant value of the mean field parameter $\alpha$=$\alpha_0$.    Fig.~\ref{Fig2}(a) summarizes the results of simulation for different values of $\alpha_0$ for a fixed value of the parameter $K$ = 10. While the remanent magnetization ($M_R$) is independent of $\alpha_0$,  the coercive field ($H_c$) decreases with the increase in $\alpha_0$. Moreover, a lower value of $\alpha_0$ generates the typical `sigmoid' type ferromagnetic hysteresis loop, whereas the magnetization  approaches to its saturation value very near to the zero field when larger values of $\alpha_0$ (e.g. 500) are considered in our simulation.   Thus, the parameter $\alpha_0$ can be tuned to study soft and hard ferromagnets, having same $M_R$. Since the parameter $K$ represents the energy loss due to all pinning processes, our simulations for a fixed $\alpha_0$ found  a decrease in both $H_c$ and $M_R$ with a lowering of the parameter $K$ [Fig.~\ref{Fig2}(b)]. We have also simulated hysteresis with a larger range of $H$ for much larger values of $K$ and the results have been shown in the Supplementary Materials.

\subsection{Effect of interface on one side of the ferromagnet}

 \begin{figure*}
\vspace{-0pt}
	\includegraphics[width=.7\textwidth] {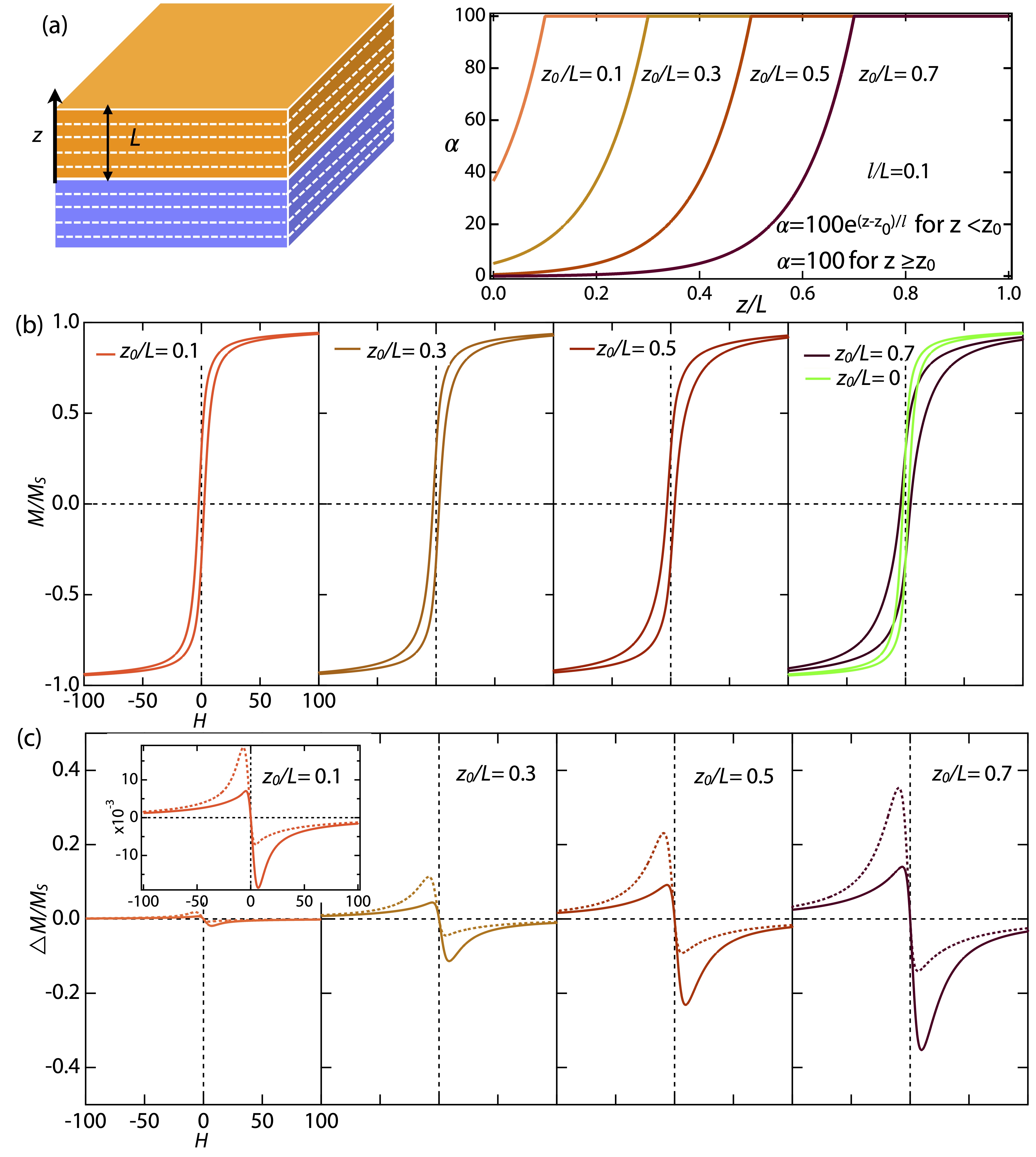}
	\caption{\label{Fig3} {Modification of ferromagnet hysteresis loop in presence of interfacial effects. (a) The mean field parameter $\alpha$ is a considered as function of the distance ($z$) from the interface:  $\alpha=\alpha_0 e^{(z-z_o)/l}$ for $z<z_0$ and  $\alpha=\alpha_0$ for $z>z_0$ (a) $\alpha$ versus $z/L$ for several values of $z_0/L$. $l/L$ is taken as 0.1.     Corresponding (b) $M/M_s$  vs. $H$ and (c)  $\Delta M/M_s$  vs. $H$  for $z_0/L$= 0.1, 0.3, 0.5 and 0.7. The dotted and solid line represents field sweep from +$H$ to -$H$, and -$H$ to +$H$, respectively. Inset in (c) is a magnified view of $M/M_s$ for  $z_0/L$= 0.1.}}
\end{figure*}

 When the ferromagnetic material is grown on a substrate or heterostructured with  another compound, the interfacial effects are very likely to affect the ferromagnetic response of the layers near the interface.  We take into account such interfacial effects  on the ferromagnetic hysteresis  by considering that the parameter $\alpha$ is no longer uniform through out all unit cells along the $z$ directions.  Since  interfacial effects very often vary  exponentially with the distance from the interface~\cite{Ohtomo:2002p378,Chakhalian:2006p244,Liu:2015p373003}, it is reasonable to assume that the $\alpha$ has the following exponential dependence with the distance from the interface
$ \alpha=\alpha_0 e^{(z-z_o)/l}$ for $z<z_0$ and  $\alpha=\alpha_0$ for $z>z_0$, where $z_0$ is a critical distance from the interface, above which interfacial effects are insignificant.  $l$ is a parameter with the dimension of the length.   The corresponding variation of $\alpha$ across the ferromagnetic material is plotted in the right panel of Fig.~\ref{Fig3}(a) as a function of the fractional distance from the interface  for different values of $z_0/L$ ($l/L$=0.1 and $\alpha_0$=100 were used for these plots).

  \begin{figure}
\vspace{-0pt}
	\includegraphics[width=.7\textwidth] {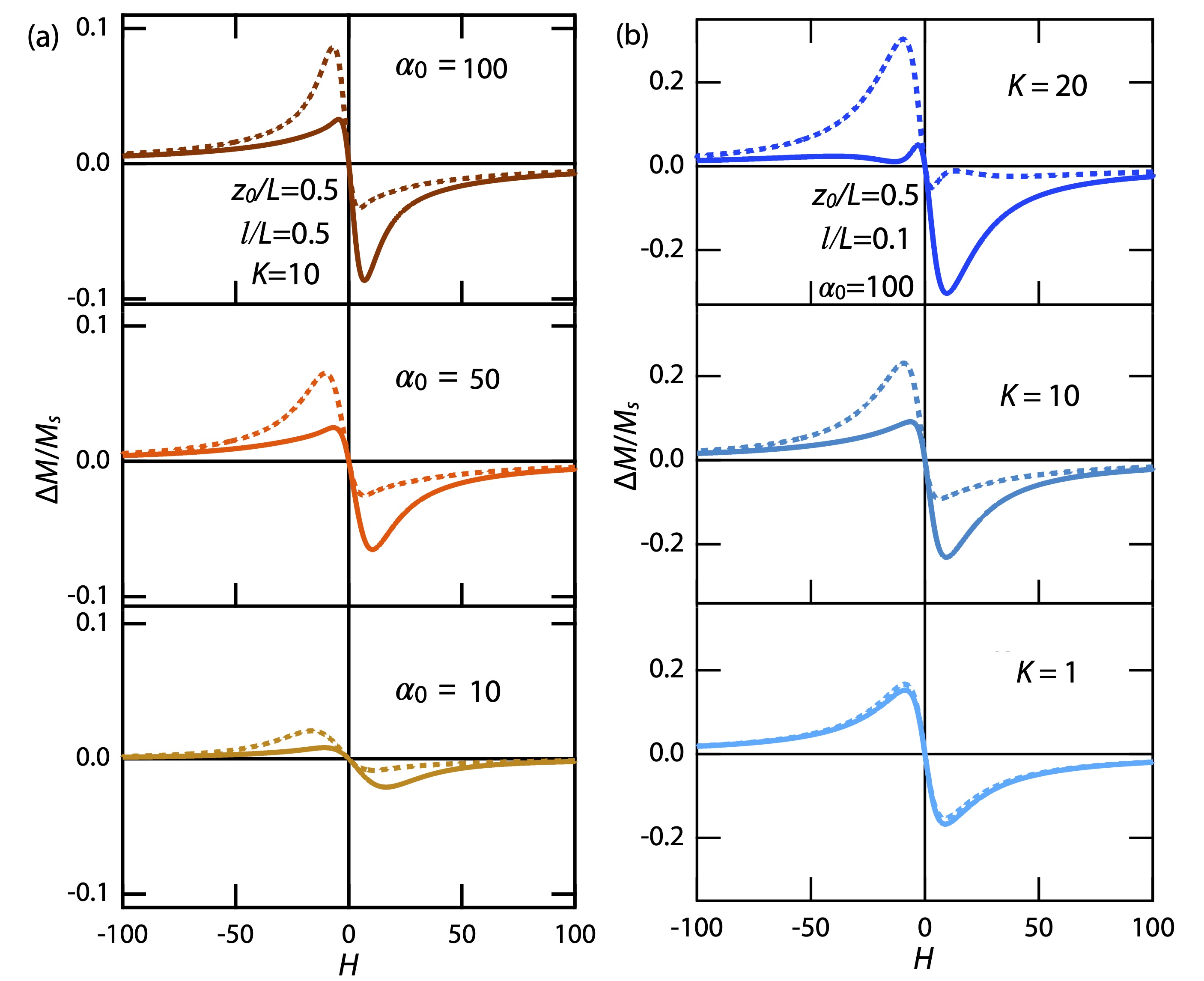}
	\caption{\label{Fig4} {Difference between magnetization with and without interfacial effects  $\Delta M/M_s$ versus $H$ (a)  for several values of $\alpha_0$ with  $z_0/L$ =0.5, $l/L$=0.5 and $K$=100; (b) for several values of $K$ with  $z_0/L$ =0.5, $l/L$=0.1 and $\alpha_0$=100.  The dotted and solid line represents field sweep from +$H$ to -$H$, and -$H$ to +$H$, respectively.}}
\end{figure}

The hysteresis curves for these distance dependent $\alpha$ with $K$= 10 have been simulated for a ferromagnetic system,  consisting of $N$=100 unit cells, similar to the previously discussed case of a uniform ferromagnet (Fig.~\ref{Fig2}).  As expected, the hysteresis loop for the $z$-dependent $\alpha$ looks quite different compared to the case of the uniform ferromagnet [see the right most panel of Fig.~\ref{Fig3}(b)].  As we  reduce the extent of interfacial effects by decreasing $z_0/L$, the hysteresis loop gradually becomes similar to the hysteresis of a uniform ferromagnet. This is further evident in Fig.~\ref{Fig3}(c), where the difference ($\Delta M$) between the magnetization response of the layer dependent case and the layer independent case have been shown for the same set of $z_0/L$.  Though the $\Delta M/M_s$ decreases with the reduction of $z_0/L$, the interfacial effect is still prominent even in case of the simulation with $z_0/L$=0.1 [see inset in Fig.~\ref{Fig3}(c)]. The hysteresis in $\Delta M/M_s$ (difference between forward and backward field sweep) also increases with $z_0/L$. Our most surprising finding is that these  $\Delta M$  would lead to an extra  contribution (=$R_a \Delta M$ in equation 2) in the Hall effect measurement,  which would resemble THE-features, reported in a variety of quantum materials in recent years~\cite{Liu:2018p122,Wang:2021p1108,Li:2019p21268,Meng:2019p3169,Yun:2018p034201,Shao:2019p182,groenendijk2020berry}.  Interestingly, the $\Delta M$, obtained from our simulations,  also varies linearly near $H$=0 - akin to  the linear variation of $\rho_{THE}$ observed experimentally~\cite{Wang:2019p1054}.

 The effect of the parameter $l$ has been explored with a fixed value of $z_0/L$=0.5 (results are shown in Supplementary Materials).  As expected, smaller value of $l$ would correspond to a stronger interfacial effect, leading to an increase of $\Delta M$/$M_s$.

We have already demonstrated in context of a uniform ferromagnet (Fig.~\ref{Fig2}(a)) that  we can explore both soft and hard ferromagnetic behaviour by changing the parameter $\alpha$= $\alpha_0$.  In order to explore how  $\alpha_0$ affects the hysteresis loop in presence of interfacial effects, we have simulated magnetization curves  considering different values of $\alpha_0$ for $z_0/L$=0.5, $l/L$ =0.5, and $K$=10. Clearly,  $\Delta M$/$M_s$ is  reduced  with the lowering of  $\alpha_0$  (Fig.~\ref{Fig4}(a)), signifying that the change in the magnetic hysteresis due to interfacial effects would be lower in a hard ferromagnet. The impact of the pinning energy scale on the interfacial effect ($\Delta M$/$M_s$) has been examined by varying the parameter $K$.  The outcomes of our simulations with lowering $K$ have been shown in Fig.~\ref{Fig4}(b)  [$z_0/L$=0.5, $l/L$ =0.1, and $\alpha_0$=100].  The hysteresis in $\Delta M$/$M_s$, which is very strong for the case of $K$=20, decreases with lowering $K$ and is completely absent for  the $K$=1 case. Similar behaviors have been observed in $\rho_\mathrm{THE}$ vs. $H$ measurements at different $T$  for SrRuO$_3$~\cite{Skoropata:2020peaaz3902,Seddon:2021p12}.

 \begin{figure*}
\vspace{-0pt}
	\includegraphics[width=.7\textwidth] {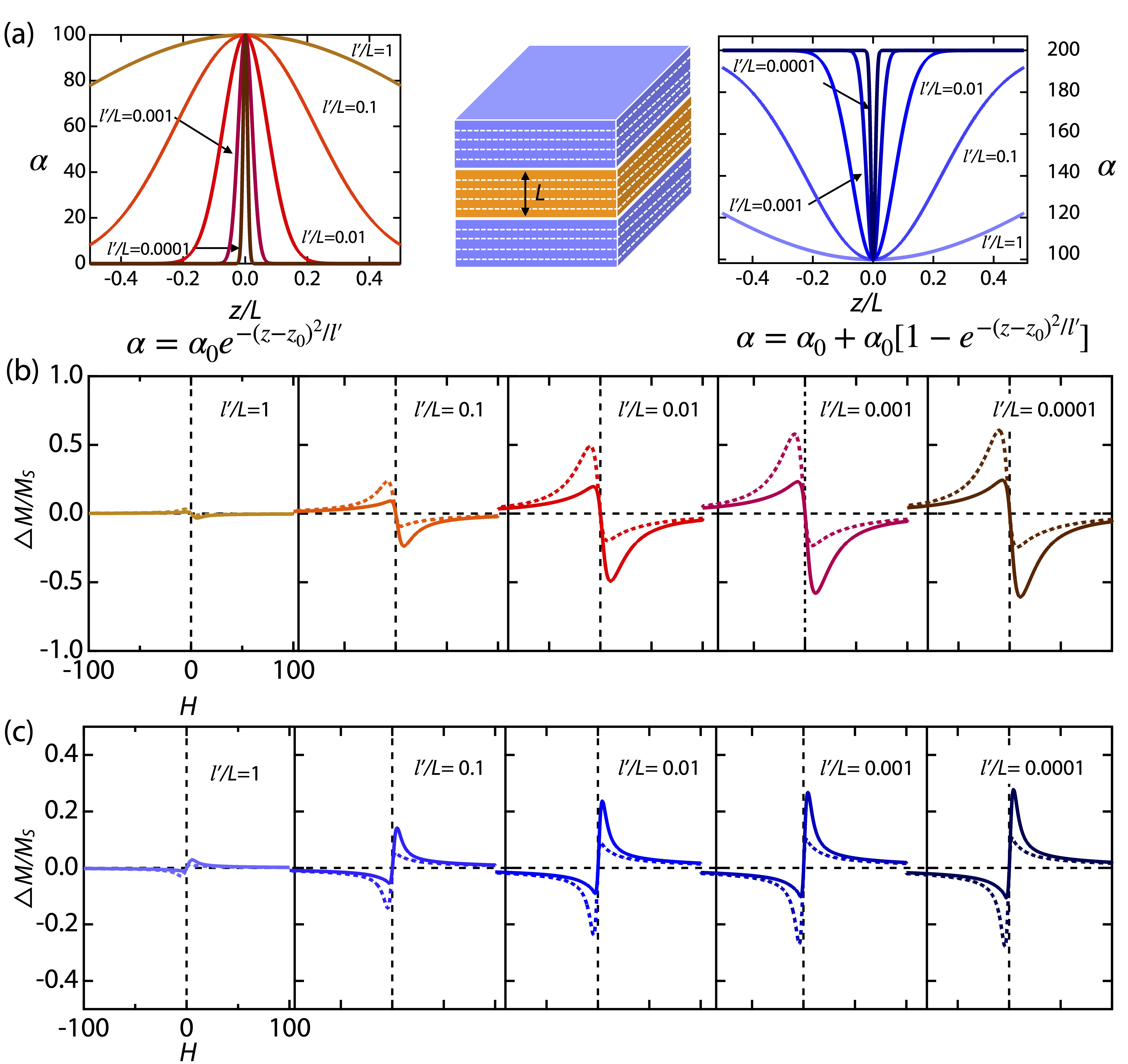}
	\caption{\label{Fig5} {Difference between magnetization with and without interfacial effects  $\Delta M/M_s$ versus $H$ for $l/L$=0.1, 0.3, 0.5 and 0.7, keeping $z_0/L$ =0.5. Inset: $\alpha$ versus $z/L$ for the corresponding values of $l$.  The dotted and solid line in (b) and (c) represents field sweep from +$H$ to -$H$, and -$H$ to +$H$, respectively. }}
\end{figure*}

\subsection{Effect of interfaces on both sides of the ferromagnet}
So far, we have investigated hysteresis  of a ferromagnet with interfacial effects on one side while the other side retains  it's bulk-like interdomain magnetic coupling parameter. The inversion symmetry is broken in such a heterostructure,  which can lead  to finite D-M interaction. Since the net D-M interaction would vanish in case of a ferromagnet having interface on both sides, the possibilities of getting  THE-like features has been also examined in SrRuO$_3$-based heterostructures with such symmetric interfaces~\cite{Wang:2019p1054,Yang:2021p014403}.  To explore the modification of hysteresis loop in such an artificial structure (see middle panel of Fig.~\ref{Fig5}(a)), we have considered the variation of $\alpha$ as a Gaussian function, where the centre of the ferromagnetic compound is taken as the $z$=0 reference point. The variations of $\alpha=\alpha_0 e^{-(z-z_0)^2/l'}$ for different values of $l'$ and $\alpha_0$=100 have been shown in left panel of Fig.~\ref{Fig5}(a). The corresponding variations of $\Delta M$/$M_s$ are shown in Fig.~\ref{Fig5}(b).  Changes in magnetization due to interfacial effects are enhanced with the lowering of the parameter $l'$. The overall behavior of $\Delta M$/$M_s$ vs. $H$  is very similar to the previously discussed ferromagnet having interfacial changes on one side. We also consider a  possible scenario where $\alpha$ is enhanced towards the interface [$\alpha=\alpha_0 + \alpha_0 (1-e^{-(z-z_0)^2/l'}$)]. The  sign of  $\Delta M$/$M_s$ is reversed compared to the previous case (Fig.~\ref{Fig5}(c)).  Thus, our  model predicts that THE-like features can be observed in a heterostructure with symmetric interfaces as well due to the layer dependent interdomain magnetic coupling even though the net D-M interaction is zero.

  We also note that the observed maximum magnitude of topological Hall resistivity in experiment is around one order of magnitude smaller compared to the maximum magnitude of anomalous Hall resistivity~\cite{Wang:2018p1087,Wang:2019p1054}. Although we have not explicitly calculated resistivity here, the relative magnitudes of the extra AHE ($\propto \Delta M$) and the AHE ($\propto M$) component in present case can be used for the comparison. With the ranges of parameters we have explored in this paper, we have found that the maximum of $\Delta M/M_s$ is around one order of magnitude smaller compared to the maximum of $M/M_s$ as well.

\section {Conclusion}
To summarize,  motivated by several recent experimental papers on SrRuO$_3$ based heterostructures that claim that additional features in AHE that are often attributed to THE may not necessarily be a result of non trivial spin textures and may arise from an inhomogeneity in the ferromagnetic structure~\cite{Kan:2018p180408,Wu:2020p220406,Kimbell:2020p054414,Wang:2020p2468,Wysocki:2020p4.054402}, we have developed a simple model to investigate changes in ferromagnetic hysteresis loop due to interfacial effects. Within the formalism of a phenomenological approach, developed by Jiles et al.~\cite{Jiles:1986p48}, we have considered that the strength of the interdomain magnetic coupling parameter within the ferromagnetic material as a function of the distance from the interface.  The resultant change in magnetization, obtained from our simulations, would lead to an additional contribution in AHE, which resembles experimental aspects of THE seen in several cases.  While we do not claim that all experimental reports of THE in oxide heterostructures can be accounted by multi-channel AHE due to  layer dependent hysteresis, we want to emphasize that such possibility must be examined before assigning these additional features in AHE to non trivial spin textures.  Though we have discussed about interfacial effects in oxide heterostructure only in this paper,  our present model is very general and can be applied for any  heterostructure consisting of elemental ferromagnets, chalcogenides, heavy fermions etc.

See the supplementary material for simulations with larger range of $H$, and the effect of the parameter $l/L$ in the case of simulations  having interfacial effects on one side.

\section{Acknowledgements}
We thank Shashank Kumar Ojha and Prithwijit Mandal for insightful comments about this paper. SM acknowledges financial supports from DST Nanomission grant (Grant No. DST/NM/NS/2018/246) and Infosys Foundation, Bangalore.\\




\begin{thebibliography}{60}%
\makeatletter
\providecommand \@ifxundefined [1]{%
 \@ifx{#1\undefined}
}%
\providecommand \@ifnum [1]{%
 \ifnum #1\expandafter \@firstoftwo
 \else \expandafter \@secondoftwo
 \fi
}%
\providecommand \@ifx [1]{%
 \ifx #1\expandafter \@firstoftwo
 \else \expandafter \@secondoftwo
 \fi
}%
\providecommand \natexlab [1]{#1}%
\providecommand \enquote  [1]{``#1''}%
\providecommand \bibnamefont  [1]{#1}%
\providecommand \bibfnamefont [1]{#1}%
\providecommand \citenamefont [1]{#1}%
\providecommand \href@noop [0]{\@secondoftwo}%
\providecommand \href [0]{\begingroup \@sanitize@url \@href}%
\providecommand \@href[1]{\@@startlink{#1}\@@href}%
\providecommand \@@href[1]{\endgroup#1\@@endlink}%
\providecommand \@sanitize@url [0]{\catcode `\\12\catcode `\$12\catcode
  `\&12\catcode `\#12\catcode `\^12\catcode `\_12\catcode `\%12\relax}%
\providecommand \@@startlink[1]{}%
\providecommand \@@endlink[0]{}%
\providecommand \url  [0]{\begingroup\@sanitize@url \@url }%
\providecommand \@url [1]{\endgroup\@href {#1}{\urlprefix }}%
\providecommand \urlprefix  [0]{URL }%
\providecommand \Eprint [0]{\href }%
\providecommand \doibase [0]{http://dx.doi.org/}%
\providecommand \selectlanguage [0]{\@gobble}%
\providecommand \bibinfo  [0]{\@secondoftwo}%
\providecommand \bibfield  [0]{\@secondoftwo}%
\providecommand \translation [1]{[#1]}%
\providecommand \BibitemOpen [0]{}%
\providecommand \bibitemStop [0]{}%
\providecommand \bibitemNoStop [0]{.\EOS\space}%
\providecommand \EOS [0]{\spacefactor3000\relax}%
\providecommand \BibitemShut  [1]{\csname bibitem#1\endcsname}%
\let\auto@bib@innerbib\@empty
\bibitem [{\citenamefont {Imada}, \citenamefont {Fujimori},\ and\ \citenamefont
  {Tokura}(1998)}]{Imada:1998p1039}%
  \BibitemOpen
  \bibfield  {author} {\bibinfo {author} {\bibfnamefont {M.}~\bibnamefont
  {Imada}}, \bibinfo {author} {\bibfnamefont {A.}~\bibnamefont {Fujimori}}, \
  and\ \bibinfo {author} {\bibfnamefont {Y.}~\bibnamefont {Tokura}},\ }\href
  {\doibase 10.1103/RevModPhys.70.1039} {\bibfield  {journal} {\bibinfo
  {journal} {Rev. Mod. Phys.}\ }\textbf {\bibinfo {volume} {70}},\ \bibinfo
  {pages} {1039} (\bibinfo {year} {1998})}\BibitemShut {NoStop}%
\bibitem [{\citenamefont {Tokura}(2006)}]{Tokura:2006p797}%
  \BibitemOpen
  \bibfield  {author} {\bibinfo {author} {\bibfnamefont {Y.}~\bibnamefont
  {Tokura}},\ }\href {\doibase 10.1088/0034-4885/69/3/r06} {\bibfield
  {journal} {\bibinfo  {journal} {Reports on Progress in Physics}\ }\textbf
  {\bibinfo {volume} {69}},\ \bibinfo {pages} {797} (\bibinfo {year}
  {2006})}\BibitemShut {NoStop}%
\bibitem [{\citenamefont {Keimer}\ \emph {et~al.}(2015)\citenamefont {Keimer},
  \citenamefont {Kivelson}, \citenamefont {Norman}, \citenamefont {Uchida},\
  and\ \citenamefont {Zaanen}}]{Keimer:2015p179}%
  \BibitemOpen
  \bibfield  {author} {\bibinfo {author} {\bibfnamefont {B.}~\bibnamefont
  {Keimer}}, \bibinfo {author} {\bibfnamefont {S.~A.}\ \bibnamefont
  {Kivelson}}, \bibinfo {author} {\bibfnamefont {M.~R.}\ \bibnamefont
  {Norman}}, \bibinfo {author} {\bibfnamefont {S.}~\bibnamefont {Uchida}}, \
  and\ \bibinfo {author} {\bibfnamefont {J.}~\bibnamefont {Zaanen}},\ }\href
  {\doibase 10.1038/nature14165} {\bibfield  {journal} {\bibinfo  {journal}
  {Nature}\ }\textbf {\bibinfo {volume} {518}},\ \bibinfo {pages} {179}
  (\bibinfo {year} {2015})}\BibitemShut {NoStop}%
\bibitem [{\citenamefont {Armitage}, \citenamefont {Fournier},\ and\
  \citenamefont {Greene}(2010)}]{Armitage:2010p2421}%
  \BibitemOpen
  \bibfield  {author} {\bibinfo {author} {\bibfnamefont {N.~P.}\ \bibnamefont
  {Armitage}}, \bibinfo {author} {\bibfnamefont {P.}~\bibnamefont {Fournier}},
  \ and\ \bibinfo {author} {\bibfnamefont {R.~L.}\ \bibnamefont {Greene}},\
  }\href {\doibase 10.1103/RevModPhys.82.2421} {\bibfield  {journal} {\bibinfo
  {journal} {Rev. Mod. Phys.}\ }\textbf {\bibinfo {volume} {82}},\ \bibinfo
  {pages} {2421} (\bibinfo {year} {2010})}\BibitemShut {NoStop}%
\bibitem [{\citenamefont {Schlom}\ \emph {et~al.}(2008)\citenamefont {Schlom},
  \citenamefont {Chen}, \citenamefont {Pan}, \citenamefont {Schmehl},\ and\
  \citenamefont {Zurbuchen}}]{Schlom:2008p2429}%
  \BibitemOpen
  \bibfield  {author} {\bibinfo {author} {\bibfnamefont {D.~G.}\ \bibnamefont
  {Schlom}}, \bibinfo {author} {\bibfnamefont {L.-Q.}\ \bibnamefont {Chen}},
  \bibinfo {author} {\bibfnamefont {X.}~\bibnamefont {Pan}}, \bibinfo {author}
  {\bibfnamefont {A.}~\bibnamefont {Schmehl}}, \ and\ \bibinfo {author}
  {\bibfnamefont {M.~A.}\ \bibnamefont {Zurbuchen}},\ }\href {\doibase
  10.1111/j.1551-2916.2008.02556.x} {\bibfield  {journal} {\bibinfo  {journal}
  {Journal of the American Ceramic Society}\ }\textbf {\bibinfo {volume}
  {91}},\ \bibinfo {pages} {2429} (\bibinfo {year} {2008})}\BibitemShut
  {NoStop}%
\bibitem [{\citenamefont {Zubko}\ \emph {et~al.}(2011)\citenamefont {Zubko},
  \citenamefont {Gariglio}, \citenamefont {Gabay}, \citenamefont {Ghosez},\
  and\ \citenamefont {Triscone}}]{Zubko:2011p141}%
  \BibitemOpen
  \bibfield  {author} {\bibinfo {author} {\bibfnamefont {P.}~\bibnamefont
  {Zubko}}, \bibinfo {author} {\bibfnamefont {S.}~\bibnamefont {Gariglio}},
  \bibinfo {author} {\bibfnamefont {M.}~\bibnamefont {Gabay}}, \bibinfo
  {author} {\bibfnamefont {P.}~\bibnamefont {Ghosez}}, \ and\ \bibinfo {author}
  {\bibfnamefont {J.-M.}\ \bibnamefont {Triscone}},\ }\href {\doibase
  10.1146/annurev-conmatphys-062910-140445} {\bibfield  {journal} {\bibinfo
  {journal} {Annual Review of Condensed Matter Physics}\ }\textbf {\bibinfo
  {volume} {2}},\ \bibinfo {pages} {141} (\bibinfo {year} {2011})},\ \Eprint
  {http://arxiv.org/abs/https://doi.org/10.1146/annurev-conmatphys-062910-140445}
  {https://doi.org/10.1146/annurev-conmatphys-062910-140445} \BibitemShut
  {NoStop}%
\bibitem [{\citenamefont {Hwang}\ \emph {et~al.}(2012)\citenamefont {Hwang},
  \citenamefont {Iwasa}, \citenamefont {Kawasaki}, \citenamefont {Keimer},
  \citenamefont {Nagaosa},\ and\ \citenamefont {Tokura}}]{Hwang:2012p103}%
  \BibitemOpen
  \bibfield  {author} {\bibinfo {author} {\bibfnamefont {H.~Y.}\ \bibnamefont
  {Hwang}}, \bibinfo {author} {\bibfnamefont {Y.}~\bibnamefont {Iwasa}},
  \bibinfo {author} {\bibfnamefont {M.}~\bibnamefont {Kawasaki}}, \bibinfo
  {author} {\bibfnamefont {B.}~\bibnamefont {Keimer}}, \bibinfo {author}
  {\bibfnamefont {N.}~\bibnamefont {Nagaosa}}, \ and\ \bibinfo {author}
  {\bibfnamefont {Y.}~\bibnamefont {Tokura}},\ }\href {\doibase
  10.1038/nmat3223} {\bibfield  {journal} {\bibinfo  {journal} {Nature Mater.}\
  }\textbf {\bibinfo {volume} {11}},\ \bibinfo {pages} {103} (\bibinfo {year}
  {2012})}\BibitemShut {NoStop}%
\bibitem [{\citenamefont {Chakhalian}\ \emph {et~al.}(2014)\citenamefont
  {Chakhalian}, \citenamefont {Freeland}, \citenamefont {Millis}, \citenamefont
  {Panagopoulos},\ and\ \citenamefont {Rondinelli}}]{chakhalian:2014p1189}%
  \BibitemOpen
  \bibfield  {author} {\bibinfo {author} {\bibfnamefont {J.}~\bibnamefont
  {Chakhalian}}, \bibinfo {author} {\bibfnamefont {J.~W.}\ \bibnamefont
  {Freeland}}, \bibinfo {author} {\bibfnamefont {A.~J.}\ \bibnamefont
  {Millis}}, \bibinfo {author} {\bibfnamefont {C.}~\bibnamefont
  {Panagopoulos}}, \ and\ \bibinfo {author} {\bibfnamefont {J.~M.}\
  \bibnamefont {Rondinelli}},\ }\href {\doibase 10.1103/RevModPhys.86.1189}
  {\bibfield  {journal} {\bibinfo  {journal} {Rev. Mod. Phys.}\ }\textbf
  {\bibinfo {volume} {86}},\ \bibinfo {pages} {1189} (\bibinfo {year}
  {2014})}\BibitemShut {NoStop}%
\bibitem [{\citenamefont {Bhattacharya}\ and\ \citenamefont
  {May}(2014)}]{Bhattacharya:2014p65}%
  \BibitemOpen
  \bibfield  {author} {\bibinfo {author} {\bibfnamefont {A.}~\bibnamefont
  {Bhattacharya}}\ and\ \bibinfo {author} {\bibfnamefont {S.~J.}\ \bibnamefont
  {May}},\ }\href {\doibase 10.1146/annurev-matsci-070813-113447} {\bibfield
  {journal} {\bibinfo  {journal} {Annual Review of Materials Research}\
  }\textbf {\bibinfo {volume} {44}},\ \bibinfo {pages} {65} (\bibinfo {year}
  {2014})},\ \Eprint
  {http://arxiv.org/abs/http://dx.doi.org/10.1146/annurev-matsci-070813-113447}
  {http://dx.doi.org/10.1146/annurev-matsci-070813-113447} \BibitemShut
  {NoStop}%
\bibitem [{\citenamefont {Stemmer}\ and\ \citenamefont
  {James~Allen}(2014)}]{Stemmer:2014p151}%
  \BibitemOpen
  \bibfield  {author} {\bibinfo {author} {\bibfnamefont {S.}~\bibnamefont
  {Stemmer}}\ and\ \bibinfo {author} {\bibfnamefont {S.}~\bibnamefont
  {James~Allen}},\ }\href {\doibase 10.1146/annurev-matsci-070813-113552}
  {\bibfield  {journal} {\bibinfo  {journal} {Annual Review of Materials
  Research}\ }\textbf {\bibinfo {volume} {44}},\ \bibinfo {pages} {151}
  (\bibinfo {year} {2014})}\BibitemShut {NoStop}%
\bibitem [{\citenamefont {Middey}\ \emph {et~al.}(2016)\citenamefont {Middey},
  \citenamefont {Chakhalian}, \citenamefont {Mahadevan}, \citenamefont
  {Freeland}, \citenamefont {Millis},\ and\ \citenamefont
  {Sarma}}]{Middey:2016p305}%
  \BibitemOpen
  \bibfield  {author} {\bibinfo {author} {\bibfnamefont {S.}~\bibnamefont
  {Middey}}, \bibinfo {author} {\bibfnamefont {J.}~\bibnamefont {Chakhalian}},
  \bibinfo {author} {\bibfnamefont {P.}~\bibnamefont {Mahadevan}}, \bibinfo
  {author} {\bibfnamefont {J.~W.}\ \bibnamefont {Freeland}}, \bibinfo {author}
  {\bibfnamefont {A.~J.}\ \bibnamefont {Millis}}, \ and\ \bibinfo {author}
  {\bibfnamefont {D.~D.}\ \bibnamefont {Sarma}},\ }\href {\doibase
  10.1146/annurev-matsci-070115-032057} {\bibfield  {journal} {\bibinfo
  {journal} {Annual Review of Materials Research}\ }\textbf {\bibinfo {volume}
  {46}},\ \bibinfo {pages} {305} (\bibinfo {year} {2016})}\BibitemShut
  {NoStop}%
\bibitem [{\citenamefont {Hellman}\ \emph {et~al.}(2017)\citenamefont
  {Hellman}, \citenamefont {Hoffmann}, \citenamefont {Tserkovnyak},
  \citenamefont {Beach}, \citenamefont {Fullerton}, \citenamefont {Leighton},
  \citenamefont {MacDonald}, \citenamefont {Ralph}, \citenamefont {Arena},
  \citenamefont {D\"urr}, \citenamefont {Fischer}, \citenamefont {Grollier},
  \citenamefont {Heremans}, \citenamefont {Jungwirth}, \citenamefont {Kimel},
  \citenamefont {Koopmans}, \citenamefont {Krivorotov}, \citenamefont {May},
  \citenamefont {Petford-Long}, \citenamefont {Rondinelli}, \citenamefont
  {Samarth}, \citenamefont {Schuller}, \citenamefont {Slavin}, \citenamefont
  {Stiles}, \citenamefont {Tchernyshyov}, \citenamefont {Thiaville},\ and\
  \citenamefont {Zink}}]{Hellman:2017p025006}%
  \BibitemOpen
  \bibfield  {author} {\bibinfo {author} {\bibfnamefont {F.}~\bibnamefont
  {Hellman}}, \bibinfo {author} {\bibfnamefont {A.}~\bibnamefont {Hoffmann}},
  \bibinfo {author} {\bibfnamefont {Y.}~\bibnamefont {Tserkovnyak}}, \bibinfo
  {author} {\bibfnamefont {G.~S.~D.}\ \bibnamefont {Beach}}, \bibinfo {author}
  {\bibfnamefont {E.~E.}\ \bibnamefont {Fullerton}}, \bibinfo {author}
  {\bibfnamefont {C.}~\bibnamefont {Leighton}}, \bibinfo {author}
  {\bibfnamefont {A.~H.}\ \bibnamefont {MacDonald}}, \bibinfo {author}
  {\bibfnamefont {D.~C.}\ \bibnamefont {Ralph}}, \bibinfo {author}
  {\bibfnamefont {D.~A.}\ \bibnamefont {Arena}}, \bibinfo {author}
  {\bibfnamefont {H.~A.}\ \bibnamefont {D\"urr}}, \bibinfo {author}
  {\bibfnamefont {P.}~\bibnamefont {Fischer}}, \bibinfo {author} {\bibfnamefont
  {J.}~\bibnamefont {Grollier}}, \bibinfo {author} {\bibfnamefont {J.~P.}\
  \bibnamefont {Heremans}}, \bibinfo {author} {\bibfnamefont {T.}~\bibnamefont
  {Jungwirth}}, \bibinfo {author} {\bibfnamefont {A.~V.}\ \bibnamefont
  {Kimel}}, \bibinfo {author} {\bibfnamefont {B.}~\bibnamefont {Koopmans}},
  \bibinfo {author} {\bibfnamefont {I.~N.}\ \bibnamefont {Krivorotov}},
  \bibinfo {author} {\bibfnamefont {S.~J.}\ \bibnamefont {May}}, \bibinfo
  {author} {\bibfnamefont {A.~K.}\ \bibnamefont {Petford-Long}}, \bibinfo
  {author} {\bibfnamefont {J.~M.}\ \bibnamefont {Rondinelli}}, \bibinfo
  {author} {\bibfnamefont {N.}~\bibnamefont {Samarth}}, \bibinfo {author}
  {\bibfnamefont {I.~K.}\ \bibnamefont {Schuller}}, \bibinfo {author}
  {\bibfnamefont {A.~N.}\ \bibnamefont {Slavin}}, \bibinfo {author}
  {\bibfnamefont {M.~D.}\ \bibnamefont {Stiles}}, \bibinfo {author}
  {\bibfnamefont {O.}~\bibnamefont {Tchernyshyov}}, \bibinfo {author}
  {\bibfnamefont {A.}~\bibnamefont {Thiaville}}, \ and\ \bibinfo {author}
  {\bibfnamefont {B.~L.}\ \bibnamefont {Zink}},\ }\href {\doibase
  10.1103/RevModPhys.89.025006} {\bibfield  {journal} {\bibinfo  {journal}
  {Rev. Mod. Phys.}\ }\textbf {\bibinfo {volume} {89}},\ \bibinfo {pages}
  {025006} (\bibinfo {year} {2017})}\BibitemShut {NoStop}%
\bibitem [{\citenamefont {Chakhalian}\ \emph {et~al.}(2006)\citenamefont
  {Chakhalian}, \citenamefont {Freeland}, \citenamefont {Srajer}, \citenamefont
  {Strempfer}, \citenamefont {Khaliullin}, \citenamefont {Cezar}, \citenamefont
  {Charlton}, \citenamefont {Dalgliesh}, \citenamefont {Bernhard},
  \citenamefont {Cristiani}, \citenamefont {Habermeier},\ and\ \citenamefont
  {Keimer}}]{Chakhalian:2006p244}%
  \BibitemOpen
  \bibfield  {author} {\bibinfo {author} {\bibfnamefont {J.}~\bibnamefont
  {Chakhalian}}, \bibinfo {author} {\bibfnamefont {J.~W.}\ \bibnamefont
  {Freeland}}, \bibinfo {author} {\bibfnamefont {G.}~\bibnamefont {Srajer}},
  \bibinfo {author} {\bibfnamefont {J.}~\bibnamefont {Strempfer}}, \bibinfo
  {author} {\bibfnamefont {G.}~\bibnamefont {Khaliullin}}, \bibinfo {author}
  {\bibfnamefont {J.~C.}\ \bibnamefont {Cezar}}, \bibinfo {author}
  {\bibfnamefont {T.}~\bibnamefont {Charlton}}, \bibinfo {author}
  {\bibfnamefont {R.}~\bibnamefont {Dalgliesh}}, \bibinfo {author}
  {\bibfnamefont {C.}~\bibnamefont {Bernhard}}, \bibinfo {author}
  {\bibfnamefont {G.}~\bibnamefont {Cristiani}}, \bibinfo {author}
  {\bibfnamefont {H.~U.}\ \bibnamefont {Habermeier}}, \ and\ \bibinfo {author}
  {\bibfnamefont {B.}~\bibnamefont {Keimer}},\ }\href
  {http://dx.doi.org/10.1038/nphys272} {\bibfield  {journal} {\bibinfo
  {journal} {Nat Phys}\ }\textbf {\bibinfo {volume} {2}},\ \bibinfo {pages}
  {244} (\bibinfo {year} {2006})}\BibitemShut {NoStop}%
\bibitem [{\citenamefont {Chakhalian}\ \emph {et~al.}(2007)\citenamefont
  {Chakhalian}, \citenamefont {Freeland}, \citenamefont {Habermeier},
  \citenamefont {Cristiani}, \citenamefont {Khaliullin}, \citenamefont {van
  Veenendaal},\ and\ \citenamefont {Keimer}}]{Chakhalian:2007p1114}%
  \BibitemOpen
  \bibfield  {author} {\bibinfo {author} {\bibfnamefont {J.}~\bibnamefont
  {Chakhalian}}, \bibinfo {author} {\bibfnamefont {J.~W.}\ \bibnamefont
  {Freeland}}, \bibinfo {author} {\bibfnamefont {H.-U.}\ \bibnamefont
  {Habermeier}}, \bibinfo {author} {\bibfnamefont {G.}~\bibnamefont
  {Cristiani}}, \bibinfo {author} {\bibfnamefont {G.}~\bibnamefont
  {Khaliullin}}, \bibinfo {author} {\bibfnamefont {M.}~\bibnamefont {van
  Veenendaal}}, \ and\ \bibinfo {author} {\bibfnamefont {B.}~\bibnamefont
  {Keimer}},\ }\href {\doibase 10.1126/science.1149338} {\bibfield  {journal}
  {\bibinfo  {journal} {Science}\ }\textbf {\bibinfo {volume} {318}},\ \bibinfo
  {pages} {1114} (\bibinfo {year} {2007})},\ \Eprint
  {http://arxiv.org/abs/http://www.sciencemag.org/content/318/5853/1114.full.pdf}
  {http://www.sciencemag.org/content/318/5853/1114.full.pdf} \BibitemShut
  {NoStop}%
\bibitem [{\citenamefont {Brinkman}\ \emph {et~al.}(2007)\citenamefont
  {Brinkman}, \citenamefont {Huijben}, \citenamefont {van Zalk}, \citenamefont
  {Huijben}, \citenamefont {Zeitler}, \citenamefont {Maan}, \citenamefont
  {van~der Wiel}, \citenamefont {Rijnders}, \citenamefont {Blank},\ and\
  \citenamefont {Hilgenkamp}}]{Brinkman:2007p493}%
  \BibitemOpen
  \bibfield  {author} {\bibinfo {author} {\bibfnamefont {A.}~\bibnamefont
  {Brinkman}}, \bibinfo {author} {\bibfnamefont {M.}~\bibnamefont {Huijben}},
  \bibinfo {author} {\bibfnamefont {M.}~\bibnamefont {van Zalk}}, \bibinfo
  {author} {\bibfnamefont {J.}~\bibnamefont {Huijben}}, \bibinfo {author}
  {\bibfnamefont {U.}~\bibnamefont {Zeitler}}, \bibinfo {author} {\bibfnamefont
  {J.~C.}\ \bibnamefont {Maan}}, \bibinfo {author} {\bibfnamefont {W.~G.}\
  \bibnamefont {van~der Wiel}}, \bibinfo {author} {\bibfnamefont
  {G.}~\bibnamefont {Rijnders}}, \bibinfo {author} {\bibfnamefont {D.~H.~A.}\
  \bibnamefont {Blank}}, \ and\ \bibinfo {author} {\bibfnamefont
  {H.}~\bibnamefont {Hilgenkamp}},\ }\href {\doibase 10.1038/nmat1931}
  {\bibfield  {journal} {\bibinfo  {journal} {Nature Materials}\ }\textbf
  {\bibinfo {volume} {6}},\ \bibinfo {pages} {493} (\bibinfo {year}
  {2007})}\BibitemShut {NoStop}%
\bibitem [{\citenamefont {May}\ \emph {et~al.}(2009)\citenamefont {May},
  \citenamefont {Ryan}, \citenamefont {Robertson}, \citenamefont {Kim},
  \citenamefont {Santos}, \citenamefont {Karapetrova}, \citenamefont
  {Zarestky}, \citenamefont {Zhai}, \citenamefont {te~Velthuis}, \citenamefont
  {Eckstein}, \citenamefont {Bader},\ and\ \citenamefont
  {Bhattacharya}}]{May:2009p892}%
  \BibitemOpen
  \bibfield  {author} {\bibinfo {author} {\bibfnamefont {S.~J.}\ \bibnamefont
  {May}}, \bibinfo {author} {\bibfnamefont {P.~J.}\ \bibnamefont {Ryan}},
  \bibinfo {author} {\bibfnamefont {J.~L.}\ \bibnamefont {Robertson}}, \bibinfo
  {author} {\bibfnamefont {J.-W.}\ \bibnamefont {Kim}}, \bibinfo {author}
  {\bibfnamefont {T.~S.}\ \bibnamefont {Santos}}, \bibinfo {author}
  {\bibfnamefont {E.}~\bibnamefont {Karapetrova}}, \bibinfo {author}
  {\bibfnamefont {J.~L.}\ \bibnamefont {Zarestky}}, \bibinfo {author}
  {\bibfnamefont {X.}~\bibnamefont {Zhai}}, \bibinfo {author} {\bibfnamefont
  {S.~G.~E.}\ \bibnamefont {te~Velthuis}}, \bibinfo {author} {\bibfnamefont
  {J.~N.}\ \bibnamefont {Eckstein}}, \bibinfo {author} {\bibfnamefont {S.~D.}\
  \bibnamefont {Bader}}, \ and\ \bibinfo {author} {\bibfnamefont
  {A.}~\bibnamefont {Bhattacharya}},\ }\href {\doibase 10.1038/nmat2557}
  {\bibfield  {journal} {\bibinfo  {journal} {Nature Materials}\ }\textbf
  {\bibinfo {volume} {8}},\ \bibinfo {pages} {892} (\bibinfo {year}
  {2009})}\BibitemShut {NoStop}%
\bibitem [{\citenamefont {Boris}\ \emph {et~al.}(2011)\citenamefont {Boris},
  \citenamefont {Matiks}, \citenamefont {Benckiser}, \citenamefont {Frano},
  \citenamefont {Popovich}, \citenamefont {Hinkov}, \citenamefont {Wochner},
  \citenamefont {Castro-Colin}, \citenamefont {Detemple}, \citenamefont
  {Malik}, \citenamefont {Bernhard}, \citenamefont {Prokscha}, \citenamefont
  {Suter}, \citenamefont {Salman}, \citenamefont {Morenzoni}, \citenamefont
  {Cristiani}, \citenamefont {Habermeier},\ and\ \citenamefont
  {Keimer}}]{Boris:2011p937}%
  \BibitemOpen
  \bibfield  {author} {\bibinfo {author} {\bibfnamefont {A.~V.}\ \bibnamefont
  {Boris}}, \bibinfo {author} {\bibfnamefont {Y.}~\bibnamefont {Matiks}},
  \bibinfo {author} {\bibfnamefont {E.}~\bibnamefont {Benckiser}}, \bibinfo
  {author} {\bibfnamefont {A.}~\bibnamefont {Frano}}, \bibinfo {author}
  {\bibfnamefont {P.}~\bibnamefont {Popovich}}, \bibinfo {author}
  {\bibfnamefont {V.}~\bibnamefont {Hinkov}}, \bibinfo {author} {\bibfnamefont
  {P.}~\bibnamefont {Wochner}}, \bibinfo {author} {\bibfnamefont
  {M.}~\bibnamefont {Castro-Colin}}, \bibinfo {author} {\bibfnamefont
  {E.}~\bibnamefont {Detemple}}, \bibinfo {author} {\bibfnamefont {V.~K.}\
  \bibnamefont {Malik}}, \bibinfo {author} {\bibfnamefont {C.}~\bibnamefont
  {Bernhard}}, \bibinfo {author} {\bibfnamefont {T.}~\bibnamefont {Prokscha}},
  \bibinfo {author} {\bibfnamefont {A.}~\bibnamefont {Suter}}, \bibinfo
  {author} {\bibfnamefont {Z.}~\bibnamefont {Salman}}, \bibinfo {author}
  {\bibfnamefont {E.}~\bibnamefont {Morenzoni}}, \bibinfo {author}
  {\bibfnamefont {G.}~\bibnamefont {Cristiani}}, \bibinfo {author}
  {\bibfnamefont {H.-U.}\ \bibnamefont {Habermeier}}, \ and\ \bibinfo {author}
  {\bibfnamefont {B.}~\bibnamefont {Keimer}},\ }\href {\doibase
  10.1126/science.1202647} {\bibfield  {journal} {\bibinfo  {journal}
  {Science}\ }\textbf {\bibinfo {volume} {332}},\ \bibinfo {pages} {937}
  (\bibinfo {year} {2011})},\ \Eprint
  {http://arxiv.org/abs/http://www.sciencemag.org/content/332/6032/937.full.pdf}
  {http://www.sciencemag.org/content/332/6032/937.full.pdf} \BibitemShut
  {NoStop}%
\bibitem [{\citenamefont {He}\ \emph {et~al.}(2012)\citenamefont {He},
  \citenamefont {Grutter}, \citenamefont {Gu}, \citenamefont {Browning},
  \citenamefont {Takamura}, \citenamefont {Kirby}, \citenamefont {Borchers},
  \citenamefont {Kim}, \citenamefont {Fitzsimmons}, \citenamefont {Zhai},
  \citenamefont {Mehta}, \citenamefont {Wong},\ and\ \citenamefont
  {Suzuki}}]{He:2012p197202}%
  \BibitemOpen
  \bibfield  {author} {\bibinfo {author} {\bibfnamefont {C.}~\bibnamefont
  {He}}, \bibinfo {author} {\bibfnamefont {A.~J.}\ \bibnamefont {Grutter}},
  \bibinfo {author} {\bibfnamefont {M.}~\bibnamefont {Gu}}, \bibinfo {author}
  {\bibfnamefont {N.~D.}\ \bibnamefont {Browning}}, \bibinfo {author}
  {\bibfnamefont {Y.}~\bibnamefont {Takamura}}, \bibinfo {author}
  {\bibfnamefont {B.~J.}\ \bibnamefont {Kirby}}, \bibinfo {author}
  {\bibfnamefont {J.~A.}\ \bibnamefont {Borchers}}, \bibinfo {author}
  {\bibfnamefont {J.~W.}\ \bibnamefont {Kim}}, \bibinfo {author} {\bibfnamefont
  {M.~R.}\ \bibnamefont {Fitzsimmons}}, \bibinfo {author} {\bibfnamefont
  {X.}~\bibnamefont {Zhai}}, \bibinfo {author} {\bibfnamefont {V.~V.}\
  \bibnamefont {Mehta}}, \bibinfo {author} {\bibfnamefont {F.~J.}\ \bibnamefont
  {Wong}}, \ and\ \bibinfo {author} {\bibfnamefont {Y.}~\bibnamefont
  {Suzuki}},\ }\href {\doibase 10.1103/PhysRevLett.109.197202} {\bibfield
  {journal} {\bibinfo  {journal} {Phys. Rev. Lett.}\ }\textbf {\bibinfo
  {volume} {109}},\ \bibinfo {pages} {197202} (\bibinfo {year}
  {2012})}\BibitemShut {NoStop}%
\bibitem [{\citenamefont {Gibert}\ \emph {et~al.}(2012)\citenamefont {Gibert},
  \citenamefont {Zubko}, \citenamefont {Scherwitzl}, \citenamefont {Iniguez},\
  and\ \citenamefont {Triscone}}]{Gibert:2012p195}%
  \BibitemOpen
  \bibfield  {author} {\bibinfo {author} {\bibfnamefont {M.}~\bibnamefont
  {Gibert}}, \bibinfo {author} {\bibfnamefont {P.}~\bibnamefont {Zubko}},
  \bibinfo {author} {\bibfnamefont {R.}~\bibnamefont {Scherwitzl}}, \bibinfo
  {author} {\bibfnamefont {J.}~\bibnamefont {Iniguez}}, \ and\ \bibinfo
  {author} {\bibfnamefont {J.-M.}\ \bibnamefont {Triscone}},\ }\href
  {http://dx.doi.org/10.1038/nmat3224} {\bibfield  {journal} {\bibinfo
  {journal} {Nat Mater}\ }\textbf {\bibinfo {volume} {11}},\ \bibinfo {pages}
  {195} (\bibinfo {year} {2012})}\BibitemShut {NoStop}%
\bibitem [{\citenamefont {Grutter}\ \emph {et~al.}(2013)\citenamefont
  {Grutter}, \citenamefont {Yang}, \citenamefont {Kirby}, \citenamefont
  {Fitzsimmons}, \citenamefont {Aguiar}, \citenamefont {Browning},
  \citenamefont {Jenkins}, \citenamefont {Arenholz}, \citenamefont {Mehta},
  \citenamefont {Alaan},\ and\ \citenamefont {Suzuki}}]{Grutter:2013p087202}%
  \BibitemOpen
  \bibfield  {author} {\bibinfo {author} {\bibfnamefont {A.~J.}\ \bibnamefont
  {Grutter}}, \bibinfo {author} {\bibfnamefont {H.}~\bibnamefont {Yang}},
  \bibinfo {author} {\bibfnamefont {B.~J.}\ \bibnamefont {Kirby}}, \bibinfo
  {author} {\bibfnamefont {M.~R.}\ \bibnamefont {Fitzsimmons}}, \bibinfo
  {author} {\bibfnamefont {J.~A.}\ \bibnamefont {Aguiar}}, \bibinfo {author}
  {\bibfnamefont {N.~D.}\ \bibnamefont {Browning}}, \bibinfo {author}
  {\bibfnamefont {C.~A.}\ \bibnamefont {Jenkins}}, \bibinfo {author}
  {\bibfnamefont {E.}~\bibnamefont {Arenholz}}, \bibinfo {author}
  {\bibfnamefont {V.~V.}\ \bibnamefont {Mehta}}, \bibinfo {author}
  {\bibfnamefont {U.~S.}\ \bibnamefont {Alaan}}, \ and\ \bibinfo {author}
  {\bibfnamefont {Y.}~\bibnamefont {Suzuki}},\ }\href {\doibase
  10.1103/PhysRevLett.111.087202} {\bibfield  {journal} {\bibinfo  {journal}
  {Phys. Rev. Lett.}\ }\textbf {\bibinfo {volume} {111}},\ \bibinfo {pages}
  {087202} (\bibinfo {year} {2013})}\BibitemShut {NoStop}%
\bibitem [{\citenamefont {Moon}\ \emph {et~al.}(2014)\citenamefont {Moon},
  \citenamefont {Colby}, \citenamefont {Wang}, \citenamefont {Karapetrova},
  \citenamefont {Schlep{\"u}tz}, \citenamefont {Fitzsimmons},\ and\
  \citenamefont {May}}]{Moon:2014p1}%
  \BibitemOpen
  \bibfield  {author} {\bibinfo {author} {\bibfnamefont {E.~J.}\ \bibnamefont
  {Moon}}, \bibinfo {author} {\bibfnamefont {R.}~\bibnamefont {Colby}},
  \bibinfo {author} {\bibfnamefont {Q.}~\bibnamefont {Wang}}, \bibinfo {author}
  {\bibfnamefont {E.}~\bibnamefont {Karapetrova}}, \bibinfo {author}
  {\bibfnamefont {C.~M.}\ \bibnamefont {Schlep{\"u}tz}}, \bibinfo {author}
  {\bibfnamefont {M.~R.}\ \bibnamefont {Fitzsimmons}}, \ and\ \bibinfo {author}
  {\bibfnamefont {S.~J.}\ \bibnamefont {May}},\ }\href {\doibase
  10.1038/ncomms6710} {\bibfield  {journal} {\bibinfo  {journal} {Nature
  Communications}\ }\textbf {\bibinfo {volume} {5}} (\bibinfo {year} {2014}),\
  10.1038/ncomms6710}\BibitemShut {NoStop}%
\bibitem [{\citenamefont {Hoffman}\ \emph {et~al.}(2016)\citenamefont
  {Hoffman}, \citenamefont {Kirby}, \citenamefont {Kwon}, \citenamefont
  {Fabbris}, \citenamefont {Meyers}, \citenamefont {Freeland}, \citenamefont
  {Martin}, \citenamefont {Heinonen}, \citenamefont {Steadman}, \citenamefont
  {Zhou}, \citenamefont {Schlep\"utz}, \citenamefont {Dean}, \citenamefont
  {te~Velthuis}, \citenamefont {Zuo},\ and\ \citenamefont
  {Bhattacharya}}]{Hoffman:2016p041038}%
  \BibitemOpen
  \bibfield  {author} {\bibinfo {author} {\bibfnamefont {J.~D.}\ \bibnamefont
  {Hoffman}}, \bibinfo {author} {\bibfnamefont {B.~J.}\ \bibnamefont {Kirby}},
  \bibinfo {author} {\bibfnamefont {J.}~\bibnamefont {Kwon}}, \bibinfo {author}
  {\bibfnamefont {G.}~\bibnamefont {Fabbris}}, \bibinfo {author} {\bibfnamefont
  {D.}~\bibnamefont {Meyers}}, \bibinfo {author} {\bibfnamefont {J.~W.}\
  \bibnamefont {Freeland}}, \bibinfo {author} {\bibfnamefont {I.}~\bibnamefont
  {Martin}}, \bibinfo {author} {\bibfnamefont {O.~G.}\ \bibnamefont
  {Heinonen}}, \bibinfo {author} {\bibfnamefont {P.}~\bibnamefont {Steadman}},
  \bibinfo {author} {\bibfnamefont {H.}~\bibnamefont {Zhou}}, \bibinfo {author}
  {\bibfnamefont {C.~M.}\ \bibnamefont {Schlep\"utz}}, \bibinfo {author}
  {\bibfnamefont {M.~P.~M.}\ \bibnamefont {Dean}}, \bibinfo {author}
  {\bibfnamefont {S.~G.~E.}\ \bibnamefont {te~Velthuis}}, \bibinfo {author}
  {\bibfnamefont {J.-M.}\ \bibnamefont {Zuo}}, \ and\ \bibinfo {author}
  {\bibfnamefont {A.}~\bibnamefont {Bhattacharya}},\ }\href {\doibase
  10.1103/PhysRevX.6.041038} {\bibfield  {journal} {\bibinfo  {journal} {Phys.
  Rev. X}\ }\textbf {\bibinfo {volume} {6}},\ \bibinfo {pages} {041038}
  (\bibinfo {year} {2016})}\BibitemShut {NoStop}%
\bibitem [{\citenamefont {Patel}\ \emph {et~al.}(2020)\citenamefont {Patel},
  \citenamefont {Meyers}, \citenamefont {Liu}, \citenamefont {Mandal},
  \citenamefont {Kareev}, \citenamefont {Shafer}, \citenamefont {Kim},
  \citenamefont {Ryan}, \citenamefont {Middey},\ and\ \citenamefont
  {Chakhalian}}]{Ranjan:2020p041113}%
  \BibitemOpen
  \bibfield  {author} {\bibinfo {author} {\bibfnamefont {R.~K.}\ \bibnamefont
  {Patel}}, \bibinfo {author} {\bibfnamefont {D.}~\bibnamefont {Meyers}},
  \bibinfo {author} {\bibfnamefont {X.}~\bibnamefont {Liu}}, \bibinfo {author}
  {\bibfnamefont {P.}~\bibnamefont {Mandal}}, \bibinfo {author} {\bibfnamefont
  {M.}~\bibnamefont {Kareev}}, \bibinfo {author} {\bibfnamefont
  {P.}~\bibnamefont {Shafer}}, \bibinfo {author} {\bibfnamefont {J.-W.}\
  \bibnamefont {Kim}}, \bibinfo {author} {\bibfnamefont {P.~J.}\ \bibnamefont
  {Ryan}}, \bibinfo {author} {\bibfnamefont {S.}~\bibnamefont {Middey}}, \ and\
  \bibinfo {author} {\bibfnamefont {J.}~\bibnamefont {Chakhalian}},\ }\href
  {\doibase 10.1063/5.0004530} {\bibfield  {journal} {\bibinfo  {journal} {APL
  Materials}\ }\textbf {\bibinfo {volume} {8}},\ \bibinfo {pages} {041113}
  (\bibinfo {year} {2020})},\ \Eprint
  {http://arxiv.org/abs/https://doi.org/10.1063/5.0004530}
  {https://doi.org/10.1063/5.0004530} \BibitemShut {NoStop}%
\bibitem [{\citenamefont {Yi}\ \emph {et~al.}(2016)\citenamefont {Yi},
  \citenamefont {Liu}, \citenamefont {Hsu}, \citenamefont {Zhang},
  \citenamefont {Choi}, \citenamefont {Kim}, \citenamefont {Chen},
  \citenamefont {Clarkson}, \citenamefont {Serrao}, \citenamefont {Arenholz},
  \citenamefont {Ryan}, \citenamefont {Xu}, \citenamefont {Birgeneau},\ and\
  \citenamefont {Ramesh}}]{Yi:2016p6397}%
  \BibitemOpen
  \bibfield  {author} {\bibinfo {author} {\bibfnamefont {D.}~\bibnamefont
  {Yi}}, \bibinfo {author} {\bibfnamefont {J.}~\bibnamefont {Liu}}, \bibinfo
  {author} {\bibfnamefont {S.-L.}\ \bibnamefont {Hsu}}, \bibinfo {author}
  {\bibfnamefont {L.}~\bibnamefont {Zhang}}, \bibinfo {author} {\bibfnamefont
  {Y.}~\bibnamefont {Choi}}, \bibinfo {author} {\bibfnamefont {J.-W.}\
  \bibnamefont {Kim}}, \bibinfo {author} {\bibfnamefont {Z.}~\bibnamefont
  {Chen}}, \bibinfo {author} {\bibfnamefont {J.~D.}\ \bibnamefont {Clarkson}},
  \bibinfo {author} {\bibfnamefont {C.~R.}\ \bibnamefont {Serrao}}, \bibinfo
  {author} {\bibfnamefont {E.}~\bibnamefont {Arenholz}}, \bibinfo {author}
  {\bibfnamefont {P.~J.}\ \bibnamefont {Ryan}}, \bibinfo {author}
  {\bibfnamefont {H.}~\bibnamefont {Xu}}, \bibinfo {author} {\bibfnamefont
  {R.~J.}\ \bibnamefont {Birgeneau}}, \ and\ \bibinfo {author} {\bibfnamefont
  {R.}~\bibnamefont {Ramesh}},\ }\href {\doibase 10.1073/pnas.1524689113}
  {\bibfield  {journal} {\bibinfo  {journal} {Proceedings of the National
  Academy of Sciences}\ }\textbf {\bibinfo {volume} {113}},\ \bibinfo {pages}
  {6397} (\bibinfo {year} {2016})}\BibitemShut {NoStop}%
\bibitem [{\citenamefont {Matsuno}\ \emph {et~al.}(2016)\citenamefont
  {Matsuno}, \citenamefont {Ogawa}, \citenamefont {Yasuda}, \citenamefont
  {Kagawa}, \citenamefont {Koshibae}, \citenamefont {Nagaosa}, \citenamefont
  {Tokura},\ and\ \citenamefont {Kawasaki}}]{Matsuno:2016pe1600304}%
  \BibitemOpen
  \bibfield  {author} {\bibinfo {author} {\bibfnamefont {J.}~\bibnamefont
  {Matsuno}}, \bibinfo {author} {\bibfnamefont {N.}~\bibnamefont {Ogawa}},
  \bibinfo {author} {\bibfnamefont {K.}~\bibnamefont {Yasuda}}, \bibinfo
  {author} {\bibfnamefont {F.}~\bibnamefont {Kagawa}}, \bibinfo {author}
  {\bibfnamefont {W.}~\bibnamefont {Koshibae}}, \bibinfo {author}
  {\bibfnamefont {N.}~\bibnamefont {Nagaosa}}, \bibinfo {author} {\bibfnamefont
  {Y.}~\bibnamefont {Tokura}}, \ and\ \bibinfo {author} {\bibfnamefont
  {M.}~\bibnamefont {Kawasaki}},\ }\href {\doibase 10.1126/sciadv.1600304}
  {\bibfield  {journal} {\bibinfo  {journal} {Science Advances}\ }\textbf
  {\bibinfo {volume} {2}},\ \bibinfo {pages} {e1600304} (\bibinfo {year}
  {2016})}\BibitemShut {NoStop}%
\bibitem [{\citenamefont {Neubauer}\ \emph {et~al.}(2009)\citenamefont
  {Neubauer}, \citenamefont {Pfleiderer}, \citenamefont {Binz}, \citenamefont
  {Rosch}, \citenamefont {Ritz}, \citenamefont {Niklowitz},\ and\ \citenamefont
  {B\"oni}}]{Neubauer:2009p186602}%
  \BibitemOpen
  \bibfield  {author} {\bibinfo {author} {\bibfnamefont {A.}~\bibnamefont
  {Neubauer}}, \bibinfo {author} {\bibfnamefont {C.}~\bibnamefont
  {Pfleiderer}}, \bibinfo {author} {\bibfnamefont {B.}~\bibnamefont {Binz}},
  \bibinfo {author} {\bibfnamefont {A.}~\bibnamefont {Rosch}}, \bibinfo
  {author} {\bibfnamefont {R.}~\bibnamefont {Ritz}}, \bibinfo {author}
  {\bibfnamefont {P.~G.}\ \bibnamefont {Niklowitz}}, \ and\ \bibinfo {author}
  {\bibfnamefont {P.}~\bibnamefont {B\"oni}},\ }\href {\doibase
  10.1103/PhysRevLett.102.186602} {\bibfield  {journal} {\bibinfo  {journal}
  {Phys. Rev. Lett.}\ }\textbf {\bibinfo {volume} {102}},\ \bibinfo {pages}
  {186602} (\bibinfo {year} {2009})}\BibitemShut {NoStop}%
\bibitem [{\citenamefont {Kanazawa}\ \emph {et~al.}(2011)\citenamefont
  {Kanazawa}, \citenamefont {Onose}, \citenamefont {Arima}, \citenamefont
  {Okuyama}, \citenamefont {Ohoyama}, \citenamefont {Wakimoto}, \citenamefont
  {Kakurai}, \citenamefont {Ishiwata},\ and\ \citenamefont
  {Tokura}}]{Kanazawa:2011p156603}%
  \BibitemOpen
  \bibfield  {author} {\bibinfo {author} {\bibfnamefont {N.}~\bibnamefont
  {Kanazawa}}, \bibinfo {author} {\bibfnamefont {Y.}~\bibnamefont {Onose}},
  \bibinfo {author} {\bibfnamefont {T.}~\bibnamefont {Arima}}, \bibinfo
  {author} {\bibfnamefont {D.}~\bibnamefont {Okuyama}}, \bibinfo {author}
  {\bibfnamefont {K.}~\bibnamefont {Ohoyama}}, \bibinfo {author} {\bibfnamefont
  {S.}~\bibnamefont {Wakimoto}}, \bibinfo {author} {\bibfnamefont
  {K.}~\bibnamefont {Kakurai}}, \bibinfo {author} {\bibfnamefont
  {S.}~\bibnamefont {Ishiwata}}, \ and\ \bibinfo {author} {\bibfnamefont
  {Y.}~\bibnamefont {Tokura}},\ }\href {\doibase
  10.1103/PhysRevLett.106.156603} {\bibfield  {journal} {\bibinfo  {journal}
  {Phys. Rev. Lett.}\ }\textbf {\bibinfo {volume} {106}},\ \bibinfo {pages}
  {156603} (\bibinfo {year} {2011})}\BibitemShut {NoStop}%
\bibitem [{\citenamefont {S{\"u}rgers}\ \emph {et~al.}(2014)\citenamefont
  {S{\"u}rgers}, \citenamefont {Fischer}, \citenamefont {Winkel},\ and\
  \citenamefont {v.~L{\"o}hneysen}}]{Surgers:2014p1}%
  \BibitemOpen
  \bibfield  {author} {\bibinfo {author} {\bibfnamefont {C.}~\bibnamefont
  {S{\"u}rgers}}, \bibinfo {author} {\bibfnamefont {G.}~\bibnamefont
  {Fischer}}, \bibinfo {author} {\bibfnamefont {P.}~\bibnamefont {Winkel}}, \
  and\ \bibinfo {author} {\bibfnamefont {H.}~\bibnamefont {v.~L{\"o}hneysen}},\
  }\href {\doibase 10.1038/ncomms4400} {\bibfield  {journal} {\bibinfo
  {journal} {Nature Communications}\ }\textbf {\bibinfo {volume} {5}} (\bibinfo
  {year} {2014}),\ 10.1038/ncomms4400}\BibitemShut {NoStop}%
\bibitem [{\citenamefont {Kurumaji}\ \emph {et~al.}(2019)\citenamefont
  {Kurumaji}, \citenamefont {Nakajima}, \citenamefont {Hirschberger},
  \citenamefont {Kikkawa}, \citenamefont {Yamasaki}, \citenamefont {Sagayama},
  \citenamefont {Nakao}, \citenamefont {Taguchi}, \citenamefont {hisa Arima},\
  and\ \citenamefont {Tokura}}]{Kurumaji:2019p6456}%
  \BibitemOpen
  \bibfield  {author} {\bibinfo {author} {\bibfnamefont {T.}~\bibnamefont
  {Kurumaji}}, \bibinfo {author} {\bibfnamefont {T.}~\bibnamefont {Nakajima}},
  \bibinfo {author} {\bibfnamefont {M.}~\bibnamefont {Hirschberger}}, \bibinfo
  {author} {\bibfnamefont {A.}~\bibnamefont {Kikkawa}}, \bibinfo {author}
  {\bibfnamefont {Y.}~\bibnamefont {Yamasaki}}, \bibinfo {author}
  {\bibfnamefont {H.}~\bibnamefont {Sagayama}}, \bibinfo {author}
  {\bibfnamefont {H.}~\bibnamefont {Nakao}}, \bibinfo {author} {\bibfnamefont
  {Y.}~\bibnamefont {Taguchi}}, \bibinfo {author} {\bibfnamefont
  {T.}~\bibnamefont {hisa Arima}}, \ and\ \bibinfo {author} {\bibfnamefont
  {Y.}~\bibnamefont {Tokura}},\ }\href {\doibase 10.1126/science.aau0968}
  {\bibfield  {journal} {\bibinfo  {journal} {Science}\ }\textbf {\bibinfo
  {volume} {365}},\ \bibinfo {pages} {914} (\bibinfo {year}
  {2019})}\BibitemShut {NoStop}%
\bibitem [{\citenamefont {Liu}\ \emph {et~al.}(2017)\citenamefont {Liu},
  \citenamefont {Zang}, \citenamefont {Ruan}, \citenamefont {Gong},
  \citenamefont {He}, \citenamefont {Ma}, \citenamefont {Xue},\ and\
  \citenamefont {Wang}}]{Liu:2017p176809}%
  \BibitemOpen
  \bibfield  {author} {\bibinfo {author} {\bibfnamefont {C.}~\bibnamefont
  {Liu}}, \bibinfo {author} {\bibfnamefont {Y.}~\bibnamefont {Zang}}, \bibinfo
  {author} {\bibfnamefont {W.}~\bibnamefont {Ruan}}, \bibinfo {author}
  {\bibfnamefont {Y.}~\bibnamefont {Gong}}, \bibinfo {author} {\bibfnamefont
  {K.}~\bibnamefont {He}}, \bibinfo {author} {\bibfnamefont {X.}~\bibnamefont
  {Ma}}, \bibinfo {author} {\bibfnamefont {Q.-K.}\ \bibnamefont {Xue}}, \ and\
  \bibinfo {author} {\bibfnamefont {Y.}~\bibnamefont {Wang}},\ }\href {\doibase
  10.1103/PhysRevLett.119.176809} {\bibfield  {journal} {\bibinfo  {journal}
  {Phys. Rev. Lett.}\ }\textbf {\bibinfo {volume} {119}},\ \bibinfo {pages}
  {176809} (\bibinfo {year} {2017})}\BibitemShut {NoStop}%
\bibitem [{\citenamefont {Ojha}\ \emph {et~al.}(2020)\citenamefont {Ojha},
  \citenamefont {Gogoi}, \citenamefont {Patidar}, \citenamefont {Patel},
  \citenamefont {Mandal}, \citenamefont {Kumar}, \citenamefont {Venkatesh},
  \citenamefont {Ganesan}, \citenamefont {Jain},\ and\ \citenamefont
  {Middey}}]{Ojha:2020p2000021}%
  \BibitemOpen
  \bibfield  {author} {\bibinfo {author} {\bibfnamefont {S.~K.}\ \bibnamefont
  {Ojha}}, \bibinfo {author} {\bibfnamefont {S.~K.}\ \bibnamefont {Gogoi}},
  \bibinfo {author} {\bibfnamefont {M.~M.}\ \bibnamefont {Patidar}}, \bibinfo
  {author} {\bibfnamefont {R.~K.}\ \bibnamefont {Patel}}, \bibinfo {author}
  {\bibfnamefont {P.}~\bibnamefont {Mandal}}, \bibinfo {author} {\bibfnamefont
  {S.}~\bibnamefont {Kumar}}, \bibinfo {author} {\bibfnamefont
  {R.}~\bibnamefont {Venkatesh}}, \bibinfo {author} {\bibfnamefont
  {V.}~\bibnamefont {Ganesan}}, \bibinfo {author} {\bibfnamefont
  {M.}~\bibnamefont {Jain}}, \ and\ \bibinfo {author} {\bibfnamefont
  {S.}~\bibnamefont {Middey}},\ }\href {\doibase 10.1002/qute.202000021}
  {\bibfield  {journal} {\bibinfo  {journal} {Advanced Quantum Technologies}\
  }\textbf {\bibinfo {volume} {3}},\ \bibinfo {pages} {2000021} (\bibinfo
  {year} {2020})}\BibitemShut {NoStop}%
\bibitem [{\citenamefont {Vistoli}\ \emph {et~al.}(2018)\citenamefont
  {Vistoli}, \citenamefont {Wang}, \citenamefont {Sander}, \citenamefont {Zhu},
  \citenamefont {Casals}, \citenamefont {Cichelero}, \citenamefont
  {Barth{\'{e}}l{\'{e}}my}, \citenamefont {Fusil}, \citenamefont {Herranz},
  \citenamefont {Valencia}, \citenamefont {Abrudan}, \citenamefont {Weschke},
  \citenamefont {Nakazawa}, \citenamefont {Kohno}, \citenamefont {Santamaria},
  \citenamefont {Wu}, \citenamefont {Garcia},\ and\ \citenamefont
  {Bibes}}]{Vistoli:2018p67}%
  \BibitemOpen
  \bibfield  {author} {\bibinfo {author} {\bibfnamefont {L.}~\bibnamefont
  {Vistoli}}, \bibinfo {author} {\bibfnamefont {W.}~\bibnamefont {Wang}},
  \bibinfo {author} {\bibfnamefont {A.}~\bibnamefont {Sander}}, \bibinfo
  {author} {\bibfnamefont {Q.}~\bibnamefont {Zhu}}, \bibinfo {author}
  {\bibfnamefont {B.}~\bibnamefont {Casals}}, \bibinfo {author} {\bibfnamefont
  {R.}~\bibnamefont {Cichelero}}, \bibinfo {author} {\bibfnamefont
  {A.}~\bibnamefont {Barth{\'{e}}l{\'{e}}my}}, \bibinfo {author} {\bibfnamefont
  {S.}~\bibnamefont {Fusil}}, \bibinfo {author} {\bibfnamefont
  {G.}~\bibnamefont {Herranz}}, \bibinfo {author} {\bibfnamefont
  {S.}~\bibnamefont {Valencia}}, \bibinfo {author} {\bibfnamefont
  {R.}~\bibnamefont {Abrudan}}, \bibinfo {author} {\bibfnamefont
  {E.}~\bibnamefont {Weschke}}, \bibinfo {author} {\bibfnamefont
  {K.}~\bibnamefont {Nakazawa}}, \bibinfo {author} {\bibfnamefont
  {H.}~\bibnamefont {Kohno}}, \bibinfo {author} {\bibfnamefont
  {J.}~\bibnamefont {Santamaria}}, \bibinfo {author} {\bibfnamefont
  {W.}~\bibnamefont {Wu}}, \bibinfo {author} {\bibfnamefont {V.}~\bibnamefont
  {Garcia}}, \ and\ \bibinfo {author} {\bibfnamefont {M.}~\bibnamefont
  {Bibes}},\ }\href {\doibase 10.1038/s41567-018-0307-5} {\bibfield  {journal}
  {\bibinfo  {journal} {Nature Physics}\ }\textbf {\bibinfo {volume} {15}},\
  \bibinfo {pages} {67} (\bibinfo {year} {2018})}\BibitemShut {NoStop}%
\bibitem [{\citenamefont {Shao}\ \emph {et~al.}(2019)\citenamefont {Shao},
  \citenamefont {Liu}, \citenamefont {Yu}, \citenamefont {Kim}, \citenamefont
  {Che}, \citenamefont {Tang}, \citenamefont {He}, \citenamefont {Tserkovnyak},
  \citenamefont {Shi},\ and\ \citenamefont {Wang}}]{Shao:2019p182}%
  \BibitemOpen
  \bibfield  {author} {\bibinfo {author} {\bibfnamefont {Q.}~\bibnamefont
  {Shao}}, \bibinfo {author} {\bibfnamefont {Y.}~\bibnamefont {Liu}}, \bibinfo
  {author} {\bibfnamefont {G.}~\bibnamefont {Yu}}, \bibinfo {author}
  {\bibfnamefont {S.~K.}\ \bibnamefont {Kim}}, \bibinfo {author} {\bibfnamefont
  {X.}~\bibnamefont {Che}}, \bibinfo {author} {\bibfnamefont {C.}~\bibnamefont
  {Tang}}, \bibinfo {author} {\bibfnamefont {Q.~L.}\ \bibnamefont {He}},
  \bibinfo {author} {\bibfnamefont {Y.}~\bibnamefont {Tserkovnyak}}, \bibinfo
  {author} {\bibfnamefont {J.}~\bibnamefont {Shi}}, \ and\ \bibinfo {author}
  {\bibfnamefont {K.~L.}\ \bibnamefont {Wang}},\ }\href {\doibase
  10.1038/s41928-019-0246-x} {\bibfield  {journal} {\bibinfo  {journal} {Nature
  Electronics}\ }\textbf {\bibinfo {volume} {2}},\ \bibinfo {pages} {182}
  (\bibinfo {year} {2019})}\BibitemShut {NoStop}%
\bibitem [{\citenamefont {Skoropata}\ \emph {et~al.}(2020)\citenamefont
  {Skoropata}, \citenamefont {Nichols}, \citenamefont {Ok}, \citenamefont
  {Chopdekar}, \citenamefont {Choi}, \citenamefont {Rastogi}, \citenamefont
  {Sohn}, \citenamefont {Gao}, \citenamefont {Yoon}, \citenamefont {Farmer},
  \citenamefont {Desautels}, \citenamefont {Choi}, \citenamefont {Haskel},
  \citenamefont {Freeland}, \citenamefont {Okamoto}, \citenamefont {Brahlek},\
  and\ \citenamefont {Lee}}]{Skoropata:2020peaaz3902}%
  \BibitemOpen
  \bibfield  {author} {\bibinfo {author} {\bibfnamefont {E.}~\bibnamefont
  {Skoropata}}, \bibinfo {author} {\bibfnamefont {J.}~\bibnamefont {Nichols}},
  \bibinfo {author} {\bibfnamefont {J.~M.}\ \bibnamefont {Ok}}, \bibinfo
  {author} {\bibfnamefont {R.~V.}\ \bibnamefont {Chopdekar}}, \bibinfo {author}
  {\bibfnamefont {E.~S.}\ \bibnamefont {Choi}}, \bibinfo {author}
  {\bibfnamefont {A.}~\bibnamefont {Rastogi}}, \bibinfo {author} {\bibfnamefont
  {C.}~\bibnamefont {Sohn}}, \bibinfo {author} {\bibfnamefont {X.}~\bibnamefont
  {Gao}}, \bibinfo {author} {\bibfnamefont {S.}~\bibnamefont {Yoon}}, \bibinfo
  {author} {\bibfnamefont {T.}~\bibnamefont {Farmer}}, \bibinfo {author}
  {\bibfnamefont {R.~D.}\ \bibnamefont {Desautels}}, \bibinfo {author}
  {\bibfnamefont {Y.}~\bibnamefont {Choi}}, \bibinfo {author} {\bibfnamefont
  {D.}~\bibnamefont {Haskel}}, \bibinfo {author} {\bibfnamefont {J.~W.}\
  \bibnamefont {Freeland}}, \bibinfo {author} {\bibfnamefont {S.}~\bibnamefont
  {Okamoto}}, \bibinfo {author} {\bibfnamefont {M.}~\bibnamefont {Brahlek}}, \
  and\ \bibinfo {author} {\bibfnamefont {H.~N.}\ \bibnamefont {Lee}},\ }\href
  {\doibase 10.1126/sciadv.aaz3902} {\bibfield  {journal} {\bibinfo  {journal}
  {Science Advances}\ }\textbf {\bibinfo {volume} {6}},\ \bibinfo {pages}
  {eaaz3902} (\bibinfo {year} {2020})}\BibitemShut {NoStop}%
\bibitem [{\citenamefont {Cao}\ \emph {et~al.}(1997)\citenamefont {Cao},
  \citenamefont {McCall}, \citenamefont {Shepard}, \citenamefont {Crow},\ and\
  \citenamefont {Guertin}}]{Cao:1997p321}%
  \BibitemOpen
  \bibfield  {author} {\bibinfo {author} {\bibfnamefont {G.}~\bibnamefont
  {Cao}}, \bibinfo {author} {\bibfnamefont {S.}~\bibnamefont {McCall}},
  \bibinfo {author} {\bibfnamefont {M.}~\bibnamefont {Shepard}}, \bibinfo
  {author} {\bibfnamefont {J.~E.}\ \bibnamefont {Crow}}, \ and\ \bibinfo
  {author} {\bibfnamefont {R.~P.}\ \bibnamefont {Guertin}},\ }\href {\doibase
  10.1103/PhysRevB.56.321} {\bibfield  {journal} {\bibinfo  {journal} {Phys.
  Rev. B}\ }\textbf {\bibinfo {volume} {56}},\ \bibinfo {pages} {321} (\bibinfo
  {year} {1997})}\BibitemShut {NoStop}%
\bibitem [{\citenamefont {Ohuchi}\ \emph {et~al.}(2018)\citenamefont {Ohuchi},
  \citenamefont {Matsuno}, \citenamefont {Ogawa}, \citenamefont {Kozuka},
  \citenamefont {Uchida}, \citenamefont {Tokura},\ and\ \citenamefont
  {Kawasaki}}]{Ohuchi:2018p1}%
  \BibitemOpen
  \bibfield  {author} {\bibinfo {author} {\bibfnamefont {Y.}~\bibnamefont
  {Ohuchi}}, \bibinfo {author} {\bibfnamefont {J.}~\bibnamefont {Matsuno}},
  \bibinfo {author} {\bibfnamefont {N.}~\bibnamefont {Ogawa}}, \bibinfo
  {author} {\bibfnamefont {Y.}~\bibnamefont {Kozuka}}, \bibinfo {author}
  {\bibfnamefont {M.}~\bibnamefont {Uchida}}, \bibinfo {author} {\bibfnamefont
  {Y.}~\bibnamefont {Tokura}}, \ and\ \bibinfo {author} {\bibfnamefont
  {M.}~\bibnamefont {Kawasaki}},\ }\href {\doibase 10.1038/s41467-017-02629-3}
  {\bibfield  {journal} {\bibinfo  {journal} {Nature Communications}\ }\textbf
  {\bibinfo {volume} {9}} (\bibinfo {year} {2018}),\
  10.1038/s41467-017-02629-3}\BibitemShut {NoStop}%
\bibitem [{\citenamefont {Wang}\ \emph {et~al.}(2018)\citenamefont {Wang},
  \citenamefont {Feng}, \citenamefont {Kim}, \citenamefont {Kim}, \citenamefont
  {Lee}, \citenamefont {Pollard}, \citenamefont {Shin}, \citenamefont {Zhou},
  \citenamefont {Peng}, \citenamefont {Lee}, \citenamefont {Meng},
  \citenamefont {Yang}, \citenamefont {Han}, \citenamefont {Kim}, \citenamefont
  {Lu},\ and\ \citenamefont {Noh}}]{Wang:2018p1087}%
  \BibitemOpen
  \bibfield  {author} {\bibinfo {author} {\bibfnamefont {L.}~\bibnamefont
  {Wang}}, \bibinfo {author} {\bibfnamefont {Q.}~\bibnamefont {Feng}}, \bibinfo
  {author} {\bibfnamefont {Y.}~\bibnamefont {Kim}}, \bibinfo {author}
  {\bibfnamefont {R.}~\bibnamefont {Kim}}, \bibinfo {author} {\bibfnamefont
  {K.~H.}\ \bibnamefont {Lee}}, \bibinfo {author} {\bibfnamefont {S.~D.}\
  \bibnamefont {Pollard}}, \bibinfo {author} {\bibfnamefont {Y.~J.}\
  \bibnamefont {Shin}}, \bibinfo {author} {\bibfnamefont {H.}~\bibnamefont
  {Zhou}}, \bibinfo {author} {\bibfnamefont {W.}~\bibnamefont {Peng}}, \bibinfo
  {author} {\bibfnamefont {D.}~\bibnamefont {Lee}}, \bibinfo {author}
  {\bibfnamefont {W.}~\bibnamefont {Meng}}, \bibinfo {author} {\bibfnamefont
  {H.}~\bibnamefont {Yang}}, \bibinfo {author} {\bibfnamefont {J.~H.}\
  \bibnamefont {Han}}, \bibinfo {author} {\bibfnamefont {M.}~\bibnamefont
  {Kim}}, \bibinfo {author} {\bibfnamefont {Q.}~\bibnamefont {Lu}}, \ and\
  \bibinfo {author} {\bibfnamefont {T.~W.}\ \bibnamefont {Noh}},\ }\href
  {\doibase 10.1038/s41563-018-0204-4} {\bibfield  {journal} {\bibinfo
  {journal} {Nature Materials}\ }\textbf {\bibinfo {volume} {17}},\ \bibinfo
  {pages} {1087} (\bibinfo {year} {2018})}\BibitemShut {NoStop}%
\bibitem [{\citenamefont {Qin}\ \emph {et~al.}(2019)\citenamefont {Qin},
  \citenamefont {Liu}, \citenamefont {Lin}, \citenamefont {Shu}, \citenamefont
  {Xie}, \citenamefont {Lim}, \citenamefont {Li}, \citenamefont {He},
  \citenamefont {Chow},\ and\ \citenamefont {Chen}}]{Qin:2019p1807008}%
  \BibitemOpen
  \bibfield  {author} {\bibinfo {author} {\bibfnamefont {Q.}~\bibnamefont
  {Qin}}, \bibinfo {author} {\bibfnamefont {L.}~\bibnamefont {Liu}}, \bibinfo
  {author} {\bibfnamefont {W.}~\bibnamefont {Lin}}, \bibinfo {author}
  {\bibfnamefont {X.}~\bibnamefont {Shu}}, \bibinfo {author} {\bibfnamefont
  {Q.}~\bibnamefont {Xie}}, \bibinfo {author} {\bibfnamefont {Z.}~\bibnamefont
  {Lim}}, \bibinfo {author} {\bibfnamefont {C.}~\bibnamefont {Li}}, \bibinfo
  {author} {\bibfnamefont {S.}~\bibnamefont {He}}, \bibinfo {author}
  {\bibfnamefont {G.~M.}\ \bibnamefont {Chow}}, \ and\ \bibinfo {author}
  {\bibfnamefont {J.}~\bibnamefont {Chen}},\ }\href {\doibase
  10.1002/adma.201807008} {\bibfield  {journal} {\bibinfo  {journal} {Advanced
  Materials}\ }\textbf {\bibinfo {volume} {31}},\ \bibinfo {pages} {1807008}
  (\bibinfo {year} {2019})}\BibitemShut {NoStop}%
\bibitem [{\citenamefont {Meng}\ \emph {et~al.}(2019)\citenamefont {Meng},
  \citenamefont {Ahmed}, \citenamefont {Ba{\'{c}}ani}, \citenamefont {Mandru},
  \citenamefont {Zhao}, \citenamefont {Bagu{\'{e}}s}, \citenamefont {Esser},
  \citenamefont {Flores}, \citenamefont {McComb}, \citenamefont {Hug},\ and\
  \citenamefont {Yang}}]{Meng:2019p3169}%
  \BibitemOpen
  \bibfield  {author} {\bibinfo {author} {\bibfnamefont {K.-Y.}\ \bibnamefont
  {Meng}}, \bibinfo {author} {\bibfnamefont {A.~S.}\ \bibnamefont {Ahmed}},
  \bibinfo {author} {\bibfnamefont {M.}~\bibnamefont {Ba{\'{c}}ani}}, \bibinfo
  {author} {\bibfnamefont {A.-O.}\ \bibnamefont {Mandru}}, \bibinfo {author}
  {\bibfnamefont {X.}~\bibnamefont {Zhao}}, \bibinfo {author} {\bibfnamefont
  {N.}~\bibnamefont {Bagu{\'{e}}s}}, \bibinfo {author} {\bibfnamefont {B.~D.}\
  \bibnamefont {Esser}}, \bibinfo {author} {\bibfnamefont {J.}~\bibnamefont
  {Flores}}, \bibinfo {author} {\bibfnamefont {D.~W.}\ \bibnamefont {McComb}},
  \bibinfo {author} {\bibfnamefont {H.~J.}\ \bibnamefont {Hug}}, \ and\
  \bibinfo {author} {\bibfnamefont {F.}~\bibnamefont {Yang}},\ }\href {\doibase
  10.1021/acs.nanolett.9b00596} {\bibfield  {journal} {\bibinfo  {journal}
  {Nano Letters}\ }\textbf {\bibinfo {volume} {19}},\ \bibinfo {pages} {3169}
  (\bibinfo {year} {2019})}\BibitemShut {NoStop}%
\bibitem [{\citenamefont {Kan}\ \emph {et~al.}(2018)\citenamefont {Kan},
  \citenamefont {Moriyama}, \citenamefont {Kobayashi},\ and\ \citenamefont
  {Shimakawa}}]{Kan:2018p180408}%
  \BibitemOpen
  \bibfield  {author} {\bibinfo {author} {\bibfnamefont {D.}~\bibnamefont
  {Kan}}, \bibinfo {author} {\bibfnamefont {T.}~\bibnamefont {Moriyama}},
  \bibinfo {author} {\bibfnamefont {K.}~\bibnamefont {Kobayashi}}, \ and\
  \bibinfo {author} {\bibfnamefont {Y.}~\bibnamefont {Shimakawa}},\ }\href
  {\doibase 10.1103/PhysRevB.98.180408} {\bibfield  {journal} {\bibinfo
  {journal} {Phys. Rev. B}\ }\textbf {\bibinfo {volume} {98}},\ \bibinfo
  {pages} {180408} (\bibinfo {year} {2018})}\BibitemShut {NoStop}%
\bibitem [{\citenamefont {Wu}\ \emph {et~al.}(2020)\citenamefont {Wu},
  \citenamefont {Wen}, \citenamefont {Fu}, \citenamefont {Wilson},
  \citenamefont {Liu}, \citenamefont {Zhang}, \citenamefont {Vasiukov},
  \citenamefont {Kareev}, \citenamefont {Pixley},\ and\ \citenamefont
  {Chakhalian}}]{Wu:2020p220406}%
  \BibitemOpen
  \bibfield  {author} {\bibinfo {author} {\bibfnamefont {L.}~\bibnamefont
  {Wu}}, \bibinfo {author} {\bibfnamefont {F.}~\bibnamefont {Wen}}, \bibinfo
  {author} {\bibfnamefont {Y.}~\bibnamefont {Fu}}, \bibinfo {author}
  {\bibfnamefont {J.~H.}\ \bibnamefont {Wilson}}, \bibinfo {author}
  {\bibfnamefont {X.}~\bibnamefont {Liu}}, \bibinfo {author} {\bibfnamefont
  {Y.}~\bibnamefont {Zhang}}, \bibinfo {author} {\bibfnamefont {D.~M.}\
  \bibnamefont {Vasiukov}}, \bibinfo {author} {\bibfnamefont {M.~S.}\
  \bibnamefont {Kareev}}, \bibinfo {author} {\bibfnamefont {J.~H.}\
  \bibnamefont {Pixley}}, \ and\ \bibinfo {author} {\bibfnamefont
  {J.}~\bibnamefont {Chakhalian}},\ }\href {\doibase
  10.1103/physrevb.102.220406} {\bibfield  {journal} {\bibinfo  {journal}
  {Physical Review B}\ }\textbf {\bibinfo {volume} {102}},\ \bibinfo {pages}
  {220406} (\bibinfo {year} {2020})}\BibitemShut {NoStop}%
\bibitem [{\citenamefont {Kimbell}\ \emph {et~al.}(2020)\citenamefont
  {Kimbell}, \citenamefont {Sass}, \citenamefont {Woltjes}, \citenamefont {Ko},
  \citenamefont {Noh}, \citenamefont {Wu},\ and\ \citenamefont
  {Robinson}}]{Kimbell:2020p054414}%
  \BibitemOpen
  \bibfield  {author} {\bibinfo {author} {\bibfnamefont {G.}~\bibnamefont
  {Kimbell}}, \bibinfo {author} {\bibfnamefont {P.~M.}\ \bibnamefont {Sass}},
  \bibinfo {author} {\bibfnamefont {B.}~\bibnamefont {Woltjes}}, \bibinfo
  {author} {\bibfnamefont {E.~K.}\ \bibnamefont {Ko}}, \bibinfo {author}
  {\bibfnamefont {T.~W.}\ \bibnamefont {Noh}}, \bibinfo {author} {\bibfnamefont
  {W.}~\bibnamefont {Wu}}, \ and\ \bibinfo {author} {\bibfnamefont {J.~W.~A.}\
  \bibnamefont {Robinson}},\ }\href {\doibase
  10.1103/PhysRevMaterials.4.054414} {\bibfield  {journal} {\bibinfo  {journal}
  {Phys. Rev. Materials}\ }\textbf {\bibinfo {volume} {4}},\ \bibinfo {pages}
  {054414} (\bibinfo {year} {2020})}\BibitemShut {NoStop}%
\bibitem [{\citenamefont {Wang}\ \emph {et~al.}(2020)\citenamefont {Wang},
  \citenamefont {Feng}, \citenamefont {Lee}, \citenamefont {Ko}, \citenamefont
  {Lu},\ and\ \citenamefont {Noh}}]{Wang:2020p2468}%
  \BibitemOpen
  \bibfield  {author} {\bibinfo {author} {\bibfnamefont {L.}~\bibnamefont
  {Wang}}, \bibinfo {author} {\bibfnamefont {Q.}~\bibnamefont {Feng}}, \bibinfo
  {author} {\bibfnamefont {H.~G.}\ \bibnamefont {Lee}}, \bibinfo {author}
  {\bibfnamefont {E.~K.}\ \bibnamefont {Ko}}, \bibinfo {author} {\bibfnamefont
  {Q.}~\bibnamefont {Lu}}, \ and\ \bibinfo {author} {\bibfnamefont {T.~W.}\
  \bibnamefont {Noh}},\ }\href {\doibase 10.1021/acs.nanolett.9b05206}
  {\bibfield  {journal} {\bibinfo  {journal} {Nano Letters}\ }\textbf {\bibinfo
  {volume} {20}},\ \bibinfo {pages} {2468} (\bibinfo {year}
  {2020})}\BibitemShut {NoStop}%
\bibitem [{\citenamefont {Wysocki}\ \emph {et~al.}(2020)\citenamefont
  {Wysocki}, \citenamefont {Yang}, \citenamefont {Gunkel}, \citenamefont
  {Dittmann}, \citenamefont {van Loosdrecht},\ and\ \citenamefont
  {Lindfors-Vrejoiu}}]{Wysocki:2020p4.054402}%
  \BibitemOpen
  \bibfield  {author} {\bibinfo {author} {\bibfnamefont {L.}~\bibnamefont
  {Wysocki}}, \bibinfo {author} {\bibfnamefont {L.}~\bibnamefont {Yang}},
  \bibinfo {author} {\bibfnamefont {F.}~\bibnamefont {Gunkel}}, \bibinfo
  {author} {\bibfnamefont {R.}~\bibnamefont {Dittmann}}, \bibinfo {author}
  {\bibfnamefont {P.~H.~M.}\ \bibnamefont {van Loosdrecht}}, \ and\ \bibinfo
  {author} {\bibfnamefont {I.}~\bibnamefont {Lindfors-Vrejoiu}},\ }\href
  {\doibase 10.1103/PhysRevMaterials.4.054402} {\bibfield  {journal} {\bibinfo
  {journal} {Phys. Rev. Materials}\ }\textbf {\bibinfo {volume} {4}},\ \bibinfo
  {pages} {054402} (\bibinfo {year} {2020})}\BibitemShut {NoStop}%
\bibitem [{\citenamefont {Groenendijk}\ \emph {et~al.}(2020)\citenamefont
  {Groenendijk}, \citenamefont {Autieri}, \citenamefont {van Thiel},
  \citenamefont {Brzezicki}, \citenamefont {Hortensius}, \citenamefont
  {Afanasiev}, \citenamefont {Gauquelin}, \citenamefont {Barone}, \citenamefont
  {van~den Bos}, \citenamefont {van Aert} \emph
  {et~al.}}]{groenendijk2020berry}%
  \BibitemOpen
  \bibfield  {author} {\bibinfo {author} {\bibfnamefont {D.~J.}\ \bibnamefont
  {Groenendijk}}, \bibinfo {author} {\bibfnamefont {C.}~\bibnamefont
  {Autieri}}, \bibinfo {author} {\bibfnamefont {T.~C.}\ \bibnamefont {van
  Thiel}}, \bibinfo {author} {\bibfnamefont {W.}~\bibnamefont {Brzezicki}},
  \bibinfo {author} {\bibfnamefont {J.}~\bibnamefont {Hortensius}}, \bibinfo
  {author} {\bibfnamefont {D.}~\bibnamefont {Afanasiev}}, \bibinfo {author}
  {\bibfnamefont {N.}~\bibnamefont {Gauquelin}}, \bibinfo {author}
  {\bibfnamefont {P.}~\bibnamefont {Barone}}, \bibinfo {author} {\bibfnamefont
  {K.}~\bibnamefont {van~den Bos}}, \bibinfo {author} {\bibfnamefont
  {S.}~\bibnamefont {van Aert}},  \emph {et~al.},\ }\href@noop {} {\bibfield
  {journal} {\bibinfo  {journal} {Physical Review Research}\ ,\ \bibinfo
  {pages} {023404}} (\bibinfo {year} {2020})}\BibitemShut {NoStop}%
\bibitem [{\citenamefont {Alberca}\ \emph {et~al.}(2011)\citenamefont
  {Alberca}, \citenamefont {Nemes}, \citenamefont {Mompean}, \citenamefont
  {Biskup}, \citenamefont {de~Andres}, \citenamefont {Munuera}, \citenamefont
  {Tornos}, \citenamefont {Leon}, \citenamefont {Hernando}, \citenamefont
  {Ferrer}, \citenamefont {Castro}, \citenamefont {Santamaria},\ and\
  \citenamefont {Garcia-Hernandez}}]{Alberca:2011p134402}%
  \BibitemOpen
  \bibfield  {author} {\bibinfo {author} {\bibfnamefont {A.}~\bibnamefont
  {Alberca}}, \bibinfo {author} {\bibfnamefont {N.~M.}\ \bibnamefont {Nemes}},
  \bibinfo {author} {\bibfnamefont {F.~J.}\ \bibnamefont {Mompean}}, \bibinfo
  {author} {\bibfnamefont {N.}~\bibnamefont {Biskup}}, \bibinfo {author}
  {\bibfnamefont {A.}~\bibnamefont {de~Andres}}, \bibinfo {author}
  {\bibfnamefont {C.}~\bibnamefont {Munuera}}, \bibinfo {author} {\bibfnamefont
  {J.}~\bibnamefont {Tornos}}, \bibinfo {author} {\bibfnamefont
  {C.}~\bibnamefont {Leon}}, \bibinfo {author} {\bibfnamefont {A.}~\bibnamefont
  {Hernando}}, \bibinfo {author} {\bibfnamefont {P.}~\bibnamefont {Ferrer}},
  \bibinfo {author} {\bibfnamefont {G.~R.}\ \bibnamefont {Castro}}, \bibinfo
  {author} {\bibfnamefont {J.}~\bibnamefont {Santamaria}}, \ and\ \bibinfo
  {author} {\bibfnamefont {M.}~\bibnamefont {Garcia-Hernandez}},\ }\href
  {\doibase 10.1103/PhysRevB.84.134402} {\bibfield  {journal} {\bibinfo
  {journal} {Phys. Rev. B}\ }\textbf {\bibinfo {volume} {84}},\ \bibinfo
  {pages} {134402} (\bibinfo {year} {2011})}\BibitemShut {NoStop}%
\bibitem [{\citenamefont {Bruno}\ \emph {et~al.}(2011)\citenamefont {Bruno},
  \citenamefont {Garcia-Barriocanal}, \citenamefont {Varela}, \citenamefont
  {Nemes}, \citenamefont {Thakur}, \citenamefont {Cezar}, \citenamefont
  {Brookes}, \citenamefont {Rivera-Calzada}, \citenamefont {Garcia-Hernandez},
  \citenamefont {Leon}, \citenamefont {Okamoto}, \citenamefont {Pennycook},\
  and\ \citenamefont {Santamaria}}]{Bruno:2011p147205}%
  \BibitemOpen
  \bibfield  {author} {\bibinfo {author} {\bibfnamefont {F.~Y.}\ \bibnamefont
  {Bruno}}, \bibinfo {author} {\bibfnamefont {J.}~\bibnamefont
  {Garcia-Barriocanal}}, \bibinfo {author} {\bibfnamefont {M.}~\bibnamefont
  {Varela}}, \bibinfo {author} {\bibfnamefont {N.~M.}\ \bibnamefont {Nemes}},
  \bibinfo {author} {\bibfnamefont {P.}~\bibnamefont {Thakur}}, \bibinfo
  {author} {\bibfnamefont {J.~C.}\ \bibnamefont {Cezar}}, \bibinfo {author}
  {\bibfnamefont {N.~B.}\ \bibnamefont {Brookes}}, \bibinfo {author}
  {\bibfnamefont {A.}~\bibnamefont {Rivera-Calzada}}, \bibinfo {author}
  {\bibfnamefont {M.}~\bibnamefont {Garcia-Hernandez}}, \bibinfo {author}
  {\bibfnamefont {C.}~\bibnamefont {Leon}}, \bibinfo {author} {\bibfnamefont
  {S.}~\bibnamefont {Okamoto}}, \bibinfo {author} {\bibfnamefont {S.~J.}\
  \bibnamefont {Pennycook}}, \ and\ \bibinfo {author} {\bibfnamefont
  {J.}~\bibnamefont {Santamaria}},\ }\href {\doibase
  10.1103/PhysRevLett.106.147205} {\bibfield  {journal} {\bibinfo  {journal}
  {Phys. Rev. Lett.}\ }\textbf {\bibinfo {volume} {106}},\ \bibinfo {pages}
  {147205} (\bibinfo {year} {2011})}\BibitemShut {NoStop}%
\bibitem [{\citenamefont {Singh}\ \emph {et~al.}(2012)\citenamefont {Singh},
  \citenamefont {Fitzsimmons}, \citenamefont {Lookman}, \citenamefont
  {Thompson}, \citenamefont {Jeen}, \citenamefont {Biswas}, \citenamefont
  {Roldan},\ and\ \citenamefont {Varela}}]{Singh:2012p077207}%
  \BibitemOpen
  \bibfield  {author} {\bibinfo {author} {\bibfnamefont {S.}~\bibnamefont
  {Singh}}, \bibinfo {author} {\bibfnamefont {M.~R.}\ \bibnamefont
  {Fitzsimmons}}, \bibinfo {author} {\bibfnamefont {T.}~\bibnamefont
  {Lookman}}, \bibinfo {author} {\bibfnamefont {J.~D.}\ \bibnamefont
  {Thompson}}, \bibinfo {author} {\bibfnamefont {H.}~\bibnamefont {Jeen}},
  \bibinfo {author} {\bibfnamefont {A.}~\bibnamefont {Biswas}}, \bibinfo
  {author} {\bibfnamefont {M.~A.}\ \bibnamefont {Roldan}}, \ and\ \bibinfo
  {author} {\bibfnamefont {M.}~\bibnamefont {Varela}},\ }\href {\doibase
  10.1103/PhysRevLett.108.077207} {\bibfield  {journal} {\bibinfo  {journal}
  {Phys. Rev. Lett.}\ }\textbf {\bibinfo {volume} {108}},\ \bibinfo {pages}
  {077207} (\bibinfo {year} {2012})}\BibitemShut {NoStop}%
\bibitem [{\citenamefont {Jiles}\ and\ \citenamefont
  {Atherton}(1986)}]{Jiles:1986p48}%
  \BibitemOpen
  \bibfield  {author} {\bibinfo {author} {\bibfnamefont {D.}~\bibnamefont
  {Jiles}}\ and\ \bibinfo {author} {\bibfnamefont {D.}~\bibnamefont
  {Atherton}},\ }\href {\doibase 10.1016/0304-8853(86)90066-1} {\bibfield
  {journal} {\bibinfo  {journal} {Journal of Magnetism and Magnetic Materials}\
  }\textbf {\bibinfo {volume} {61}},\ \bibinfo {pages} {48} (\bibinfo {year}
  {1986})}\BibitemShut {NoStop}%
\bibitem [{\citenamefont {Liu}\ \emph {et~al.}(2018)\citenamefont {Liu},
  \citenamefont {Burigu}, \citenamefont {Zhang}, \citenamefont {Jafri},
  \citenamefont {Ma}, \citenamefont {Liu}, \citenamefont {Wang},\ and\
  \citenamefont {Wu}}]{Liu:2018p122}%
  \BibitemOpen
  \bibfield  {author} {\bibinfo {author} {\bibfnamefont {Z.}~\bibnamefont
  {Liu}}, \bibinfo {author} {\bibfnamefont {A.}~\bibnamefont {Burigu}},
  \bibinfo {author} {\bibfnamefont {Y.}~\bibnamefont {Zhang}}, \bibinfo
  {author} {\bibfnamefont {H.~M.}\ \bibnamefont {Jafri}}, \bibinfo {author}
  {\bibfnamefont {X.}~\bibnamefont {Ma}}, \bibinfo {author} {\bibfnamefont
  {E.}~\bibnamefont {Liu}}, \bibinfo {author} {\bibfnamefont {W.}~\bibnamefont
  {Wang}}, \ and\ \bibinfo {author} {\bibfnamefont {G.}~\bibnamefont {Wu}},\
  }\href@noop {} {\bibfield  {journal} {\bibinfo  {journal} {Scripta
  Materialia}\ }\textbf {\bibinfo {volume} {143}},\ \bibinfo {pages} {122}
  (\bibinfo {year} {2018})}\BibitemShut {NoStop}%
\bibitem [{\citenamefont {Wang}\ \emph {et~al.}(2021)\citenamefont {Wang},
  \citenamefont {Zhao}, \citenamefont {Wang}, \citenamefont {Daniels},
  \citenamefont {Chang}, \citenamefont {Zang}, \citenamefont {Xiao},\ and\
  \citenamefont {Wu}}]{Wang:2021p1108}%
  \BibitemOpen
  \bibfield  {author} {\bibinfo {author} {\bibfnamefont {W.}~\bibnamefont
  {Wang}}, \bibinfo {author} {\bibfnamefont {Y.-F.}\ \bibnamefont {Zhao}},
  \bibinfo {author} {\bibfnamefont {F.}~\bibnamefont {Wang}}, \bibinfo {author}
  {\bibfnamefont {M.~W.}\ \bibnamefont {Daniels}}, \bibinfo {author}
  {\bibfnamefont {C.-Z.}\ \bibnamefont {Chang}}, \bibinfo {author}
  {\bibfnamefont {J.}~\bibnamefont {Zang}}, \bibinfo {author} {\bibfnamefont
  {D.}~\bibnamefont {Xiao}}, \ and\ \bibinfo {author} {\bibfnamefont
  {W.}~\bibnamefont {Wu}},\ }\href@noop {} {\bibfield  {journal} {\bibinfo
  {journal} {Nano letters}\ }\textbf {\bibinfo {volume} {21}},\ \bibinfo
  {pages} {1108} (\bibinfo {year} {2021})}\BibitemShut {NoStop}%
\bibitem [{\citenamefont {Li}\ \emph {et~al.}(2019)\citenamefont {Li},
  \citenamefont {Zhang}, \citenamefont {Zhang}, \citenamefont {Li},
  \citenamefont {Yang}, \citenamefont {Deng}, \citenamefont {Gu},\ and\
  \citenamefont {Wu}}]{Li:2019p21268}%
  \BibitemOpen
  \bibfield  {author} {\bibinfo {author} {\bibfnamefont {Y.}~\bibnamefont
  {Li}}, \bibinfo {author} {\bibfnamefont {L.}~\bibnamefont {Zhang}}, \bibinfo
  {author} {\bibfnamefont {Q.}~\bibnamefont {Zhang}}, \bibinfo {author}
  {\bibfnamefont {C.}~\bibnamefont {Li}}, \bibinfo {author} {\bibfnamefont
  {T.}~\bibnamefont {Yang}}, \bibinfo {author} {\bibfnamefont {Y.}~\bibnamefont
  {Deng}}, \bibinfo {author} {\bibfnamefont {L.}~\bibnamefont {Gu}}, \ and\
  \bibinfo {author} {\bibfnamefont {D.}~\bibnamefont {Wu}},\ }\href@noop {}
  {\bibfield  {journal} {\bibinfo  {journal} {ACS applied materials \&
  interfaces}\ }\textbf {\bibinfo {volume} {11}},\ \bibinfo {pages} {21268}
  (\bibinfo {year} {2019})}\BibitemShut {NoStop}%
\bibitem [{\citenamefont {Yun}\ \emph {et~al.}(2018)\citenamefont {Yun},
  \citenamefont {Ma}, \citenamefont {Su}, \citenamefont {Xing}, \citenamefont
  {Chen}, \citenamefont {Yao}, \citenamefont {Cai}, \citenamefont {Yuan},\ and\
  \citenamefont {Han}}]{Yun:2018p034201}%
  \BibitemOpen
  \bibfield  {author} {\bibinfo {author} {\bibfnamefont {Y.}~\bibnamefont
  {Yun}}, \bibinfo {author} {\bibfnamefont {Y.}~\bibnamefont {Ma}}, \bibinfo
  {author} {\bibfnamefont {T.}~\bibnamefont {Su}}, \bibinfo {author}
  {\bibfnamefont {W.}~\bibnamefont {Xing}}, \bibinfo {author} {\bibfnamefont
  {Y.}~\bibnamefont {Chen}}, \bibinfo {author} {\bibfnamefont {Y.}~\bibnamefont
  {Yao}}, \bibinfo {author} {\bibfnamefont {R.}~\bibnamefont {Cai}}, \bibinfo
  {author} {\bibfnamefont {W.}~\bibnamefont {Yuan}}, \ and\ \bibinfo {author}
  {\bibfnamefont {W.}~\bibnamefont {Han}},\ }\href {\doibase
  10.1103/PhysRevMaterials.2.034201} {\bibfield  {journal} {\bibinfo  {journal}
  {Phys. Rev. Materials}\ }\textbf {\bibinfo {volume} {2}},\ \bibinfo {pages}
  {034201} (\bibinfo {year} {2018})}\BibitemShut {NoStop}%
\bibitem [{\citenamefont {Langevin}(1905)}]{Langevin:1905p203}%
  \BibitemOpen
  \bibfield  {author} {\bibinfo {author} {\bibfnamefont {P.}~\bibnamefont
  {Langevin}},\ }\href@noop {} {\bibfield  {journal} {\bibinfo  {journal} {Ann.
  chim. et phys.}\ ,\ \bibinfo {pages} {203}} (\bibinfo {year}
  {1905})}\BibitemShut {NoStop}%
\bibitem [{\citenamefont {Li}, \citenamefont {Qiao},\ and\ \citenamefont
  {Tang}(2017)}]{Li:2017numerical}%
  \BibitemOpen
  \bibfield  {author} {\bibinfo {author} {\bibfnamefont {Z.}~\bibnamefont
  {Li}}, \bibinfo {author} {\bibfnamefont {Z.}~\bibnamefont {Qiao}}, \ and\
  \bibinfo {author} {\bibfnamefont {T.}~\bibnamefont {Tang}},\ }\href@noop {}
  {\emph {\bibinfo {title} {Numerical solution of differential equations:
  introduction to finite difference and finite element methods}}}\ (\bibinfo
  {publisher} {Cambridge University Press},\ \bibinfo {year}
  {2017})\BibitemShut {NoStop}%
\bibitem [{\citenamefont {Ohtomo}\ \emph {et~al.}(2002)\citenamefont {Ohtomo},
  \citenamefont {Muller}, \citenamefont {Grazul},\ and\ \citenamefont
  {Hwang}}]{Ohtomo:2002p378}%
  \BibitemOpen
  \bibfield  {author} {\bibinfo {author} {\bibfnamefont {A.}~\bibnamefont
  {Ohtomo}}, \bibinfo {author} {\bibfnamefont {D.~A.}\ \bibnamefont {Muller}},
  \bibinfo {author} {\bibfnamefont {J.~L.}\ \bibnamefont {Grazul}}, \ and\
  \bibinfo {author} {\bibfnamefont {H.~Y.}\ \bibnamefont {Hwang}},\ }\href
  {\doibase 10.1038/nature00977} {\bibfield  {journal} {\bibinfo  {journal}
  {Nature}\ }\textbf {\bibinfo {volume} {419}},\ \bibinfo {pages} {378}
  (\bibinfo {year} {2002})}\BibitemShut {NoStop}%
\bibitem [{\citenamefont {Liu}\ and\ \citenamefont
  {Ke}(2015)}]{Liu:2015p373003}%
  \BibitemOpen
  \bibfield  {author} {\bibinfo {author} {\bibfnamefont {Y.}~\bibnamefont
  {Liu}}\ and\ \bibinfo {author} {\bibfnamefont {X.}~\bibnamefont {Ke}},\
  }\href {\doibase 10.1002/chin.201547213} {\bibfield  {journal} {\bibinfo
  {journal} {{ChemInform}}\ }\textbf {\bibinfo {volume} {46}},\ \bibinfo
  {pages} {373003} (\bibinfo {year} {2015})}\BibitemShut {NoStop}%
\bibitem [{\citenamefont {Wang}\ \emph {et~al.}(2019)\citenamefont {Wang},
  \citenamefont {Daniels}, \citenamefont {Liao}, \citenamefont {Zhao},
  \citenamefont {Wang}, \citenamefont {Koster}, \citenamefont {Rijnders},
  \citenamefont {Chang}, \citenamefont {Xiao},\ and\ \citenamefont
  {Wu}}]{Wang:2019p1054}%
  \BibitemOpen
  \bibfield  {author} {\bibinfo {author} {\bibfnamefont {W.}~\bibnamefont
  {Wang}}, \bibinfo {author} {\bibfnamefont {M.~W.}\ \bibnamefont {Daniels}},
  \bibinfo {author} {\bibfnamefont {Z.}~\bibnamefont {Liao}}, \bibinfo {author}
  {\bibfnamefont {Y.}~\bibnamefont {Zhao}}, \bibinfo {author} {\bibfnamefont
  {J.}~\bibnamefont {Wang}}, \bibinfo {author} {\bibfnamefont {G.}~\bibnamefont
  {Koster}}, \bibinfo {author} {\bibfnamefont {G.}~\bibnamefont {Rijnders}},
  \bibinfo {author} {\bibfnamefont {C.-Z.}\ \bibnamefont {Chang}}, \bibinfo
  {author} {\bibfnamefont {D.}~\bibnamefont {Xiao}}, \ and\ \bibinfo {author}
  {\bibfnamefont {W.}~\bibnamefont {Wu}},\ }\href@noop {} {\bibfield  {journal}
  {\bibinfo  {journal} {Nature materials}\ }\textbf {\bibinfo {volume} {18}},\
  \bibinfo {pages} {1054} (\bibinfo {year} {2019})}\BibitemShut {NoStop}%
\bibitem [{\citenamefont {Seddon}\ \emph {et~al.}(2021)\citenamefont {Seddon},
  \citenamefont {Dogaru}, \citenamefont {Holt}, \citenamefont {Rusu},
  \citenamefont {Peters}, \citenamefont {Sanchez},\ and\ \citenamefont
  {Alexe}}]{Seddon:2021p12}%
  \BibitemOpen
  \bibfield  {author} {\bibinfo {author} {\bibfnamefont {S.~D.}\ \bibnamefont
  {Seddon}}, \bibinfo {author} {\bibfnamefont {D.~E.}\ \bibnamefont {Dogaru}},
  \bibinfo {author} {\bibfnamefont {S.~J.~R.}\ \bibnamefont {Holt}}, \bibinfo
  {author} {\bibfnamefont {D.}~\bibnamefont {Rusu}}, \bibinfo {author}
  {\bibfnamefont {J.~J.~P.}\ \bibnamefont {Peters}}, \bibinfo {author}
  {\bibfnamefont {A.~M.}\ \bibnamefont {Sanchez}}, \ and\ \bibinfo {author}
  {\bibfnamefont {M.}~\bibnamefont {Alexe}},\ }\href {\doibase
  10.1038/s41467-021-22165-5} {\bibfield  {journal} {\bibinfo  {journal}
  {Nature Communications}\ }\textbf {\bibinfo {volume} {12}} (\bibinfo {year}
  {2021}),\ 10.1038/s41467-021-22165-5}\BibitemShut {NoStop}%
\bibitem [{\citenamefont {Yang}\ \emph {et~al.}(2021)\citenamefont {Yang},
  \citenamefont {Wysocki}, \citenamefont {Sch\"opf}, \citenamefont {Jin},
  \citenamefont {Kov\'acs}, \citenamefont {Gunkel}, \citenamefont {Dittmann},
  \citenamefont {van Loosdrecht},\ and\ \citenamefont
  {Lindfors-Vrejoiu}}]{Yang:2021p014403}%
  \BibitemOpen
  \bibfield  {author} {\bibinfo {author} {\bibfnamefont {L.}~\bibnamefont
  {Yang}}, \bibinfo {author} {\bibfnamefont {L.}~\bibnamefont {Wysocki}},
  \bibinfo {author} {\bibfnamefont {J.}~\bibnamefont {Sch\"opf}}, \bibinfo
  {author} {\bibfnamefont {L.}~\bibnamefont {Jin}}, \bibinfo {author}
  {\bibfnamefont {A.}~\bibnamefont {Kov\'acs}}, \bibinfo {author}
  {\bibfnamefont {F.}~\bibnamefont {Gunkel}}, \bibinfo {author} {\bibfnamefont
  {R.}~\bibnamefont {Dittmann}}, \bibinfo {author} {\bibfnamefont {P.~H.~M.}\
  \bibnamefont {van Loosdrecht}}, \ and\ \bibinfo {author} {\bibfnamefont
  {I.}~\bibnamefont {Lindfors-Vrejoiu}},\ }\href {\doibase
  10.1103/PhysRevMaterials.5.014403} {\bibfield  {journal} {\bibinfo  {journal}
  {Phys. Rev. Materials}\ }\textbf {\bibinfo {volume} {5}},\ \bibinfo {pages}
  {014403} (\bibinfo {year} {2021})}\BibitemShut {NoStop}%
\end{thebibliography}
%

\setcounter{figure}{0}
\makeatletter
\renewcommand{\thefigure}{S\@arabic\c@figure}
\makeatother
\section*{Supplementary Materials}

\subsection*{Hysteresis simulation with a larger range of $H$}
\begin{figure}[h!]
   \vspace{-0pt}
	\includegraphics[width=.5\textwidth] {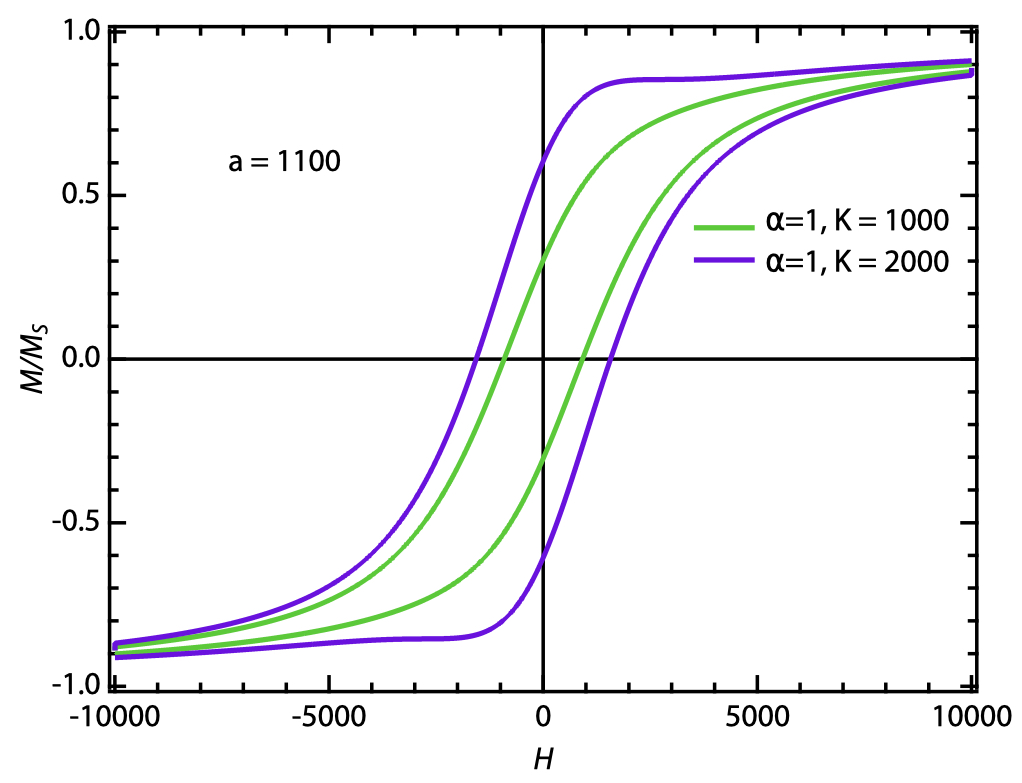}
	\caption{\label{Fig6} { Simulated hysteresis curves of a uniform ferromagnet (wi
thout any interface) for different values of $K$ keeping $\alpha_0$=1, for higher values of $H$.}}
	\end{figure}

We have simulated hysteresis for a uniformly magnetized ferromagnet with larger range of $H$ by considering $a$=1100 (Fig.~\ref{Fig6}).

\section*{Role of the parameter $l/L$}
\begin{figure*}[h!]
\vspace{-0pt}
	\includegraphics[width=.9\textwidth] {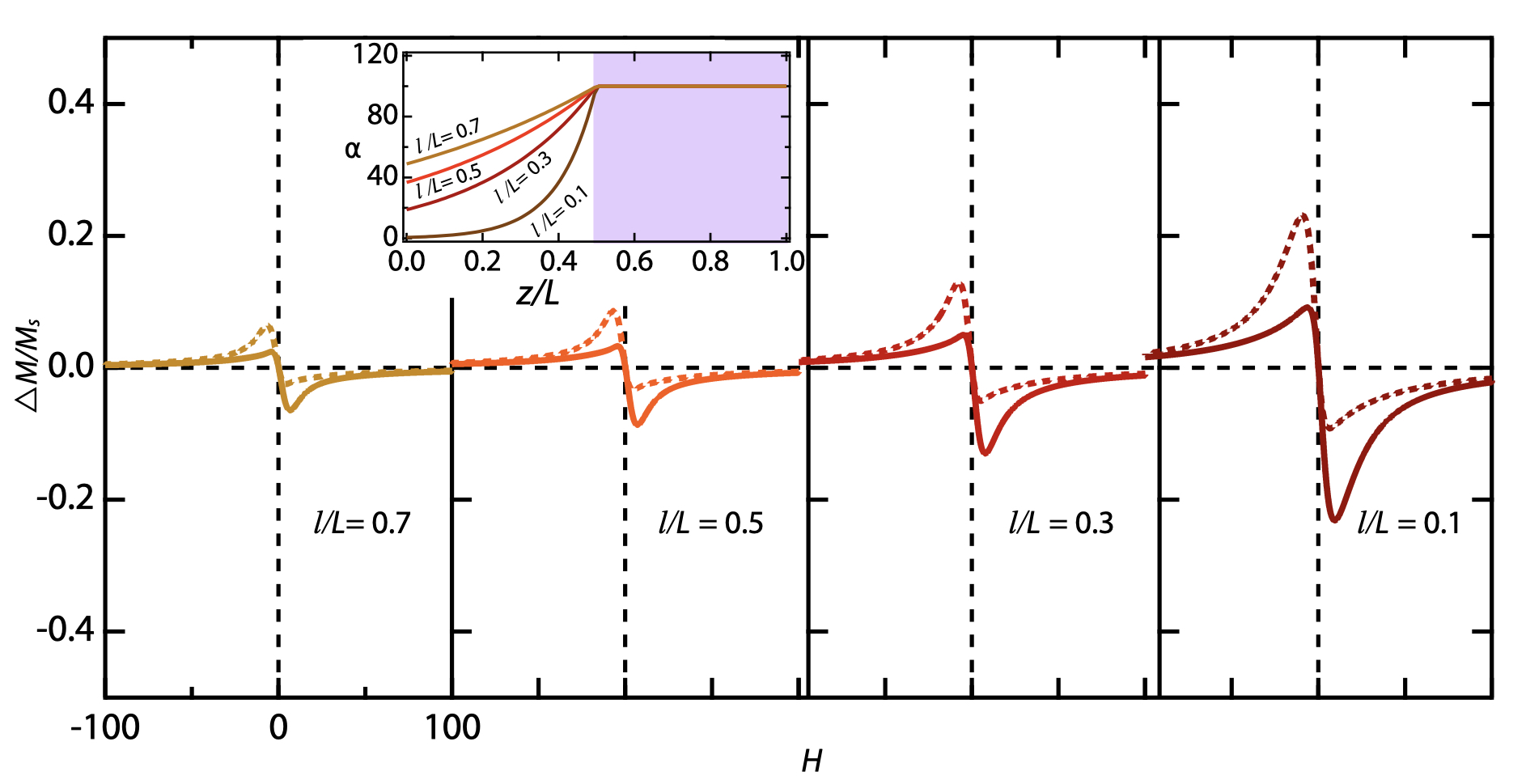}
	\caption{\label{Figsup2} {Modification of ferromagnet hysteresis loop in presence of interfacial effects. The mean field parameter $\alpha$ is considered as a function of the distance ($z$) from the interface:  $\alpha=\alpha_0 e^{(z-z_o)/l}$ for $z<z_0$ and  $\alpha=\alpha_0$ for $z>z_0$,   $\Delta M/M_s$  vs. $H$  for $l/L$= 0.7, 0.5, 0.3 and 0.1, inset: $\alpha$ versus $z/L$ for several values of $l/L$. $z_0/L$ is taken as 0.5. }}
\end{figure*}

For the case of the ferromagnet having interfacial effects on one side, we have plotted the $\Delta M/M_s$ curves for various values of $z_0/L$ in Fig. 3 of main text. Another parameter that affects the exponential nature of $\alpha$ is $l$. We expect to have similar trends in the plots of $\Delta M/M_s$ if we consider different values of $l$. The results of the simulations with different values of $l/L$ are shown in (Fig.~\ref{Figsup2}).

\end{document}